\documentclass[final,times,twocolumn,sort&compress]{elsarticle}
\usepackage{graphicx}
\usepackage{amssymb}

\journal{Journal of Magnetism and Magnetic Materials}

\begin{document}

\begin{frontmatter}
\title{Anomalous spin frustration enforced by a magnetoelastic coupling in the mixed-spin Ising model on decorated planar lattices\tnoteref{grant}}       
\tnotetext[grant]{This work was financially supported by the grant of The Ministry of Education, Science, Research and Sport of the Slovak Republic under the contract No. VEGA 1/0331/15 and by the grant of the Slovak Research and Development Agency under the contract 
No. APVV-14-0073.}
\author[upjs]{Jozef Stre\v{c}ka\corref{cor1}}
\cortext[cor1]{Corresponding author.}
\ead{jozef.strecka@upjs.sk}
\author[upjs]{Mat\'{u}\v{s} Rebi\v{c}}
\author[ufla]{Onofre Rojas} 
\author[ufla]{Sergio Martins de Souza} 
\address[upjs]{Department of Theoretical Physics and Astrophysics, Faculty of Science, P. J. \v{S}af\'{a}rik University, Park Angelinum 9, 040 01 Ko\v{s}ice, Slovakia}
\address[ufla]{Departamento de F\'isica, Universidade Federal de Lavras, CP 3037, 37200-000, Lavras-MG, Brazil}

\begin{abstract}
The mixed spin-1/2 and spin-$S$ Ising model on a decorated planar lattice accounting for lattice vibrations of decorating atoms is treated by making use of the canonical coordinate transformation, the decoration-iteration transformation, and the harmonic approximation. It is shown that the magnetoelastic coupling gives rise to an effective single-ion anisotropy and three-site four-spin interaction, which are responsible for the anomalous spin frustration of the decorating spins in virtue of a competition with the equilibrium nearest-neighbor exchange interaction between the nodal and decorating spins. The ground-state and finite-temperature phase diagrams are constructed for the particular case of the mixed spin-1/2 and spin-1 Ising model on a decorated square lattice for which thermal dependencies of the spontaneous magnetization and specific heat are also examined in detail. It is evidenced that a sufficiently strong magnetoelastic coupling leads 
to a peculiar coexistence of the antiferromagnetic long-range order of the nodal spins with the disorder of the decorating spins within the frustrated antiferromagnetic phase, which may also exhibit double reentrant phase transitions. The investigated model displays a variety of temperature dependencies of the total specific heat, which may involve in its magnetic part one or two logarithmic divergences apart from one or two additional round maxima superimposed on a standard thermal dependence of the lattice part of the specific heat. 
\end{abstract}


\end{frontmatter}

\section{Introduction}
\label{intro}

A complete thermodynamic description of magnetic, vibrational and elastic properties of insulating solid-state materials remains a long-standing problem of particular research interest, because considerable computational difficulties arise when magnetic and lattice degrees of freedom are coupled together through a magnetoelastic interaction. Owing to this fact, the magnetoelastic coupling is often completely disregarded in order to preserve a capability of treating magnetic and lattice degrees of freedom of magnetic solids independently of each other. A substantial progress in this research area has been recently made by Balcerzak and co-workers when developing a phenomenological theory based on a self-consistent variational approach, which combines different approximations for individual subsystems as for instance a mean-field approximation for a magnetic subsystem, Debey approximation for a lattice subsystem, etc. \cite{bal17,sza18,bal18} 

It is notorious that the leading-order interaction term between localized spins of insulating magnetic solids is an indirect superexchange coupling, which according to the Kramers-Anderson mechanism basically depends on an overlap of atomic wave functions \cite{kra34,and50}. The superexchange coupling (further referred to as the exchange coupling) thus strongly depends on an instantaneous distance between the magnetic atoms, which are subject to a perpetual temperature-dependent lattice vibrations. Hence, it follows that the lattice vibrations of magnetic atoms can fundamentally influence a magnetic spin ordering and vice versa. This effect might be especially marked in a close vicinity of phase transitions connected with a breakdown of a spontaneous long-range order \cite{bea62,ber76,hen87,mas00,bou01,pli10}. 

The main goal of the present work is to examine an effect of the magnetoelastic coupling on a full thermodynamics of the mixed spin-1/2 and spin-$S$ Ising model on decorated planar lattices, which are constituted by nodal atoms placed at rigid lattice positions and decorating atoms capable of lattice vibrations treated within the harmonic approximation. To this end, we will generalize the calculation procedure developed in our previous work for the analogous spin-1/2 Ising model on decorated planar lattices \cite{str12}. Interestingly, the same local canonical transformation can be applied to decouple magnetic and lattice degrees of freedom \cite{ent73}, but the relevant  decoupling gives rise to an effective three-site four-spin interaction and a shift of uniaxial single-ion anisotropy in contrast to the previous case with an effective next-nearest-neighbor interaction \cite{str12}. The magnetoelastic coupling may thus enforce a remarkable spin frustration of the decorating atoms, which has been comprehensively studied in the mixed-spin Ising model with the three-site four-spin interaction on decorated planar lattices \cite{jas16,stu17} with the help of exact mapping transformation method 
\cite{fis59,syo72,roj09,str10,str20,roj11}.  
 
The organization of this paper is as follows. The investigated mixed-spin Ising model is defined in Section \ref{model}, where the basic steps of the calculation procedure are also explained. The most interesting results for the ground-state and finite-temperature phase diagrams, the spontaneous magnetization and the specific heat are presented in Section \ref{result}. Finally, some conclusions and future outlooks are mentioned in Section \ref{conclusion}. 

\section{Model and method}
\label{model}

\begin{figure}
\begin{center}
\includegraphics[width=0.3\textwidth]{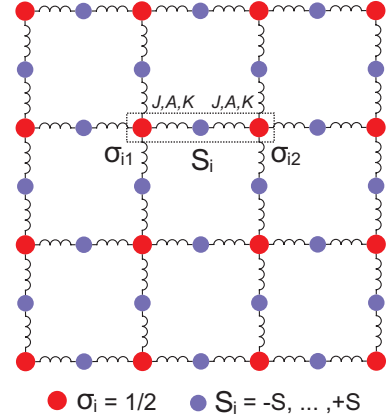}
\end{center}
\vspace{-0.6cm}
\caption{A cross-section from a decorated square lattice. Nodal lattice sites (red circles) are occupied by the spin-1/2 atoms 
(${\sigma}_{i}= \pm 1/2$), while decorating lattice sites (blue circles) are occupied by the spin-$S$ atoms ($S_i = -S, -S+1, \dots, S$) being subject to lattice vibrations.}
\label{fig1}       
\end{figure}

Let us consider a two-dimensional decorated lattice as schematically illustrated in Fig.~\ref{fig1} on the particular example of a decorated square lattice, the nodal sites of which are occupied by the spin-$1/2$ atoms and the decorating sites of which are occupied by the spin-$S$ ($S \geq 1$) atoms. It is assumed that the spin-1/2 nodal atoms are placed at rigid lattice positions in contrast to the spin-$S$ decorating atoms, which may oscillate around their equilibrium lattice positions. This approximation is justified because the relaxation of the nodal atoms from their equilibrium lattice positions would cost a greater amount of the elastic energy in comparison with the one of the decorating atoms due to a deformation of greater number of lattice bonds. Under these assumptions, the total Hamiltonian of the mixed-spin Ising model on decorated planar lattices can be defined as a sum over bond Hamiltonians ${\hat{\cal H}}_{i}$ involving all interaction terms of the $i$th decorating atom
\begin{eqnarray}
{\hat{\cal H}}\!\!\!&=&\!\!\!\sum_{i=1}^{Nq/2}{\hat{\cal H}}_{i}=\sum_{i=1}^{Nq/2}({\hat{\cal H}}_{i}^{m}+{\hat{\cal H}}_{i}^{e}),
\label{ha}
\end{eqnarray}
which are further split into the magnetoelastic part ${\hat{\cal H}}_{i}^{m}$ and the pure elastic part ${\hat{\cal H}}_{i}^{e}$ ($N$ labels the total number of the spin-1/2 nodal atoms and $q$ is their coordination number). The magnetoelastic part of the bond Hamiltonian ${\hat{\cal H}}_{i}^{m}$ takes into account the uniaxial single-ion anisotropy $D$ acting on the decorating spin $S_i$ as well as the exchange interaction between the decorating spin $S_{i}$ and its two nearest-neighbor nodal spins $\sigma_{i1}$ and $\sigma_{i2}$
\begin{eqnarray}
{\hat{\cal H}}_{i}^{m}\!\!\!&=&\!\!\!-\left(J-A\hat{\rho}_{i}\right)S_{i}{\sigma}_{i1}
                                     -\left(J+A\hat{\rho}_{i}\right)S_{i}{\sigma}_{i2}-DS_{i}^2,
\label{ham}
\end{eqnarray}
which depends on an instantaneous distance between the relevant spins through the local coordinate operator $\hat{\rho}_{i}$ assigned to a displacement of the $i$th decorating atom from its equilibrium lattice position placed at a midpoint in between its two nearest-neighbor nodal atoms. The exchange constant $J$ marks a size of the nearest-neighbor interaction between the decorating and nodal spins on assumption that the decorating atom takes its equilibrium position (i.e. $\rho_{i} = 0$), whereas the magnetoelastic coupling constant $A$ determines an increase (decrease) of the nearest-neighbor interaction owing to a contraction (elongation) of the respective atomic distance. 

The purely elastic part of the bond Hamiltonian ${\hat{\cal H}}_{i}^{e}$ incorporates the kinetic energy of the $i$th decorating atom with the mass $M$ and the elastic energy penalty, which is in the harmonic approximation connected to a square of the displacement operator of the $i$th decorating atom from its equilibrium lattice position 
\begin{eqnarray}
{\hat{\cal H}}_{i}^{e}\!\!\!&=&\!\!\!\frac{{\hat{p}_{i}^{2}}}{2M}+K\hat{\rho}_{i}^2.
\label{hal}
\end{eqnarray}
The spring stiffness constant $K$ emerging in the elastic part of the bond Hamiltonian (\ref{hal}) characterizes a vibrational energy of the decorating atoms and it can be alternatively viewed as a bare elastic constant of two harmonic springs attached to each decorating atom. 

It is evident from Eq. (\ref{ham}) that the magnetic and lattice degrees of freedom are coupled together through the magnetoelastic constant $A$, which usually makes solution of the respective Ising models refined with a magnetoelastic coupling more complex. However, the magnetoelastic interaction can be decoupled through a local canonical coordinate transformation \cite{str12,ent73} 
\begin{eqnarray}
\hat{\rho}_{i}\!\!\!&=&\!\!\!\hat{\rho}_{i}^{\prime}-\frac{A}{2K}S_{i}\left({\sigma}_{i1}-{\sigma}_{i2}\right),
\label{cct}
\end{eqnarray}
which eliminates from the magnetoelastic part of bond Hamiltonian (\ref{ham}) dependence on a displacement operator   
\begin{eqnarray}
{\hat{\cal H}}_{i}^{m \prime}\!\!\!&=&\!\!\!-J S_{i}\left({\sigma}_{i1}+{\sigma}_{i2}\right)
+J^{\prime}S_{i}^{2}{\sigma}_{i1}{\sigma}_{i2}-D^{\prime}S_{i}^{2}.
\label{hamt}
\end{eqnarray}
After implementation of the canonical coordinate transformation (\ref{cct}) the magnetoelastic part of the bond Hamiltonian (\ref{hamt}) actually involves the effective three-site four-spin interaction $J^{\prime}=A^{2}/2K$ and the rescaled uniaxial single-ion anisotropy $D^{\prime}=D+A^{2}/8K$ besides the equilibrium nearest-neighbor bilinear interaction $J$. Recently, it has been demonstrated that the mixed-spin Ising models on a decorated square lattice with the nearest-neighbor bilinear interaction, three-site four-spin interaction and unixial single-ion anisotropy may exhibit striking frustrated states due to competing effects arising from the three-site four-spin interaction \cite{jas16,stu17}.  On the other hand, the elastic part of the bond Hamiltonian (\ref{hal}) remains invariant under the canonical coordinate transformation (\ref{cct})  
\begin{eqnarray}
{\hat{\cal H}}_{i}^{e \prime}\!\!\!&=&\!\!\!\frac{\hat{p}_{i}^{\prime 2}}{2M}+K\hat{\rho}_{i}^{\prime 2}
\label{halt}
\end{eqnarray}
and it can be subsequently brought into the diagonal form 
\begin{eqnarray}
{\hat{\cal H}}_{i}^{e \prime}=\hbar\omega\left(\hat{b}_{i}^{+}\hat{b}_{i}^{-}+\frac{1}{2}\right)
\label{haltac}
\end{eqnarray}
by introducing the annihilation and creation bosonic operators with the angular frequency of normal-mode oscillations $\omega=\sqrt{2K/M}$ \begin{eqnarray}
\hat{b}_{i}^{+}=\sqrt{\frac{M\omega}{2\hbar}}\left(\hat{\rho}_{i}^{\prime}-\frac{i}{M\omega}\hat{p}_{i}^{\prime}\right), \, \, \,
\hat{b}_{i}^{-}=\sqrt{\frac{M\omega}{2\hbar}}\left(\hat{\rho}_{i}^{\prime}+\frac{i}{M\omega}\hat{p}_{i}^{\prime}\right). \nonumber
\label{ac}
\end{eqnarray}

It should be pointed out that all bond Hamiltonians ${\hat{\cal H}}_{i}^{m \prime}$ and ${\hat{\cal H}}_{i}^{e \prime}$ given by Eqs. (\ref{hamt}) and (\ref{haltac}) commute with each other and hence, the partition function of the mixed-spin Ising model on a decorated planar lattice can be partially factorized into a product of separate elastic and magnetic bond partition functions
\begin{eqnarray}
{\cal Z}\!\!\!&=&\!\!\!\sum_{\{{\sigma}_{i}\}}\prod_{i=1}^{Nq/2}\left[{\mbox{Tr}}_{i}\exp\left(-\beta{\hat{\cal H}}_{i}^{e \prime}\right)\right]\left[\sum_{S_{i}=-S}^{+S}\exp\left(-\beta{\hat{\cal H}}_{i}^{m \prime}\right)\right] \nonumber \\ \!\!\!&=&\!\!\!\sum_{\{{\sigma}_{i}\}}\prod_{i=1}^{Nq/2}{\cal Z}_{i}^{e}{\cal Z}_{i}^{m}.
\label{z}
\end{eqnarray}
In above, $\beta =1/(k_{\rm B}T)$, $k_{\rm B}$ is the Boltzmann's constant, $T$ is the absolute temperature, the symbol ${\mbox{Tr}}_{i}$ denotes a trace over the lattice degrees of freedom of the $i$th decorating atom, the summation $\sum_{\{{\sigma}_{i}\}}$ is carried out  over all states of the nodal spins and the last summation $\sum_{S_{i}}$ runs over all states of the $i$th decorating spin $S$. The elastic part of the bond partition function ${\cal Z}_{i}^{e}$ straightforwardly follows from a diagonal form of the elastic part of the bond Hamiltonian (\ref{haltac}) with regard to a trace invariance 
\begin{eqnarray}
{\cal Z}_{i}^{e}\!\!\!&=&\!\!\!{\mbox{Tr}}_{i}\exp\left(-\beta{\hat{\cal H}}_{i}^{e \prime}\right)
={\left[2\sinh\left(\frac{\beta\hbar\omega}{2}\right)\right]}^{-1},
\label{zl}
\end{eqnarray}
while the magnetic part of the bond partition function ${\cal Z}_{i}^{m}$ can be replaced with the generalized decoration-iteration transformation \cite{fis59,syo72,roj09,str10,str20,roj11}
\begin{eqnarray}
{\cal Z}_{i}^{m}\!\!\!\!\!&=&\!\!\!\!\!\!\sum_{n=-S}^{+S}\!\!\exp\left(\beta D^{\prime}n^{2}\!\!-\!\beta J^{\prime}n^{2}{\sigma}_{i1}{\sigma}_{i2}\right) \cosh\left[\beta Jn\left({\sigma}_{i1}\!\!+\!{\sigma}_{i2}\right)\right]\nonumber \\ &=&\!\!\!\! R_{0}\exp\left(\beta R_{1}{\sigma}_{i1}{\sigma}_{i2}\right).
\label{dit}
\end{eqnarray}
The mapping parameters $R_{0}$ and $R_{1}$ are unambiguously given by the self-consistency condition of the decoration-iteration transformation, which must hold for all four available states of two nodal spins ${\sigma}_{i1}$ and ${\sigma}_{i2}$ providing from the algebraic transformation (\ref{dit}) just two independent equations. Hence, it follows that the transformation parameters $R_{0}$ and $R_{1}$must obey the formulas
\begin{eqnarray}
R_{0}\!\!\!&=&\!\!\!(V_{1}V_{2})^\frac{1}{2}, \qquad \beta R_{1} = 2\ln\left(\frac{V_{1}}{V_{2}}\right),
\label{mp}
\end{eqnarray}
where for the abbreviation purposes we have introduced the following notation for expressions
\begin{eqnarray}
V_{1}\!\!\!&=&\!\!\!\sum_{n=-S}^{+S}\exp\left(\beta Dn^2\right)\cosh\left(\beta Jn\right), \nonumber \\
V_{2}\!\!\!&=&\!\!\!\sum_{n=-S}^{+S}\exp\left[\beta n^2\left(D+\frac{A^2}{4K}\right)\right].
\label{v}
\end{eqnarray}
A substitution of the elastic part of the partition function (\ref{zl}) and the decoration-iteration transformation (\ref{dit}) with properly chosen mapping parameters (\ref{mp})-(\ref{v}) into the formula (\ref{z}) affords a rigorous mapping relationship between the partition function of the mixed spin-1/2 and spin-$S$ Ising model on a decorated planar lattice prone to lattice vibrations of the decorating atoms and, respectively, the partition function of the spin-1/2 Ising model on a corresponding undecorated lattice with the effective temperature-dependent nearest-neighbor interaction $R_1$
\begin{eqnarray}
{\cal Z}\!\!\!&=&\!\!\!\left[\frac{V_{1}V_{2}}{4\sinh^{2}\left(\frac{\beta\hbar\omega}{2}\right)}\right]^{Nq/4}
{\cal Z}_{\rm IM}\left(\beta R_{1}\right),
\label{mr}
\end{eqnarray}
The partition function of the spin-1/2 Ising model was rigorously calculated for several planar lattices \cite{ons44,hou50,dom60,mco73,lav99,str15} and thus, the mapping relationship (\ref{mr}) can be utilized for obtaining exact results for the mixed-spin Ising model on decorated planar lattices. The critical points of the mixed spin-1/2 and spin-$S$ Ising model on decorated planar lattices can be for instance easily obtained from a comparison of the effective temperature-dependent nearest-neighbor coupling $\beta R_1$ with the relevant critical temperature of the spin-1/2 Ising model on corresponding undecorated planar lattice, e.g. $\beta_{C} |R_1| = 2 \ln (1 + \sqrt{2})$ for the particular case of a square lattice \cite{ons44,hou50,dom60,mco73,lav99,str15}. 

By exploiting the rigorous mapping relationship (\ref{mr}) let us also calculate basic magnetic and thermodynamic response functions of the mixed-spin Ising model on a decorated planar lattice. The overall internal energy can be calculated using the formula ${\cal U}_{tot}=-{\partial(\ln{\cal Z})}/{\partial \beta}$, which yields after straightforward manipulations the final expression ${\cal U}_{tot} = {\cal U}_{latt} + {\cal U}_{mag}$ with
\begin{eqnarray}
{\cal U}_{latt} \!\!\!&=&\!\!\!  \frac{Nq}{4}\hbar\omega\coth\left(\frac{\beta\hbar\omega}{2}\right), \label{ul} \\ 
{\cal U}_{mag} \!\!\!&=&\!\!\! -\frac{Nq}{4}\left[\frac{V_{1}^{\prime}}{V_{1}}+\frac{V_{2}^{\prime}}{V_{2}}
                  +4{\varepsilon}_{\rm IM}\left(\frac{V_{1}^{\prime}}{V_{1}}-\frac{V_{2}^{\prime}}{V_{2}}\right)\right].
\label{um}
\end{eqnarray}
The quantity $\varepsilon_{\rm IM}\equiv\langle{\sigma}_{i1}{\sigma}_{i2}\rangle$ denotes the nearest-neighbor pair correlation function of the corresponding spin-1/2 Ising model on undecorated lattice \cite{dom60} and the parameters $V_{1}^{\prime}$, $V_{2}^{\prime}$ are defined as follows   
\begin{eqnarray}
V_{1}^{\prime}\!\!\!&=&\!\!\!\sum_{n=-S}^{+S}Dn^2\exp\left(\beta Dn^2\right)\cosh\left(\beta Jn\right) \nonumber \\ &+&\!\! \sum_{n=-S}^{+S}\exp\left(\beta Dn^2\right)Jn\sinh\left(\beta Jn\right), \nonumber \\
V_{2}^{\prime}\!\!\!&=&\!\!\!\sum_{n=-S}^{+S}n^2\left(D+\frac{A^2}{4K}\right)\exp\left[\beta n^2\left(D+\frac{A^2}{4K}\right)\right].
\label{v22}
\end{eqnarray} 
The former part ${\cal U}_{latt}$ given by Eq. (\ref{ul}) determines the pure vibrational contribution to the overall internal energy, while the latter part ${\cal U}_{mag}$ given by Eq. (\ref{um}) determines the respective magnetoelastic contribution. Consequently, it is also possible to calculate the lattice contribution $C_{latt}=\partial{\cal U}_{latt}/\partial T$ as well as the magnetoelastic contribution $C_{mag}=\partial{\cal U}_{mag}/\partial T$ to the overall heat capacity $C_{tot}=C_{latt} + C_{mag}$. 

The spontaneous uniform and staggered magnetization of the nodal spins $m_{A}$ and $m_A^s$ can be computed with the help of exact mapping theorems \cite{bar88,kha90,bar91,bar95}, according to which an ensemble average $\langle \cdots \rangle$ in the mixed-spin Ising model on a decorated lattice equals to an ensemble average $\langle \cdots \rangle_{\rm IM}$ in the corresponding spin-1/2 Ising model on an undecorated rigid lattice 
\begin{eqnarray}
m_{A} \!\!\!&\equiv&\!\!\! \frac{1}{2} \langle\sigma_{i1} + \sigma_{i2}\rangle=
             \frac{1}{2} \langle\sigma_{i1} + \sigma_{i2}\rangle_{\rm IM} \equiv m_{\rm IM} (\beta R_{1}), \nonumber \\
m_{A}^s \!\!\!&\equiv&\!\!\! \frac{1}{2} \langle\sigma_{i1} - \sigma_{i2}\rangle=
             \frac{1}{2} \langle\sigma_{i1} - \sigma_{i2}\rangle_{\rm IM} \equiv m_{\rm IM}^s (\beta R_{1}).
\label{ma1}
\end{eqnarray}
Exact results for the spontaneous uniform and staggered magnetization of the nodal spins $m_{A}$ and $m_A^s$ of the mixed-spin Ising model on a decorated lattice thus directly follow according to Eq. (\ref{ma1}) from the spontaneous uniform and staggered magnetization of the corresponding spin-1/2 Ising model on the undecorated lattice \cite{lin92}. There is no principal difference between the uniform and staggered magnetization of the spin-1/2 Ising model on loose-packed planar lattices for which they represent relevant order parameters in case of ferromagnetic and antiferromagnetic nearest-neighbor interaction, respectively. As a matter of fact, the uniform magnetization of the ferromagnetic spin-1/2 Ising model on a square lattice ($R_1>0$) directly equals to the staggered magnetization of the antiferromagnetic spin-1/2 Ising model on a square lattice ($R_1<0$) \cite{yan52} 
\begin{eqnarray}
m_{\rm IM} (R_1>0) = m_{\rm IM}^s (R_1<0) = \frac{1}{2}{\left[1-\frac{1}{{\sinh}^4\left(\frac{\beta R_{1}}{2}\right)}\right]}^{\frac{1}{8}}.
\label{ma2}
\end{eqnarray}
Furthermore, the spontaneous magnetization of the decorating spins $m_{B}$ can be calculated by employing the generalized Callen-Suzuki identity \cite{cal63,suz65,bal02} 
\begin{eqnarray}
m_{B} \equiv \langle S_{i} \rangle = \biggl\langle \frac{\displaystyle\sum\limits_{S_{i}=-S}^{+S}S_{i}\exp\left(-\beta {{\cal H}_{i}^{m \prime}}\right)}{\displaystyle\sum\limits_{S_{i}=-S}^{+S}\exp\left(-\beta {{\cal H}_{i}^{m \prime}}\right)}\biggr\rangle.
\label{cs}
\end{eqnarray}
A substitution of the relations (\ref{hamt}) and (\ref{dit}) into the exact spin identity (\ref{cs}) affords the following expression for the spontaneous magnetization of the decorating spins
\begin{eqnarray}
m_{B}\!\!\!&=&\!\!\!2m_{A} \frac{\displaystyle\sum\limits_{n=-S}^{+S} n\exp\left(\beta Dn^{2}\right) \sinh\left(\beta Jn\right)}{\displaystyle\sum\limits_{n=-S}^{+S} \exp\left(\beta Dn^{2}\right) \cosh\left(\beta Jn\right)},
\label{mb}
\end{eqnarray} 
which is expressed in terms of the uniform spontaneous magnetization of the nodal spins (\ref{ma1})-(\ref{ma2}). 

Finally, it is also worth calculating a mean displacement and standard deviation of the decorating atoms from their equilibrium lattice positions, which will help us to delimit a parameter space where the harmonic approximation is applicable. According to the coordinate transformation (\ref{cct}), the mean displacement of the decorating atoms can be related to a difference between two nearest-neighbor pair correlation functions and the mean displacement in a new coordinate system   
\begin{eqnarray}
\langle\hat{\rho}_{i}\rangle\!\!\!&=&\!\!\!\langle\hat{\rho}_{i}^{\prime}\rangle-\frac{A}{2K}\left(\langle S_{i}{\sigma}_{i1}\rangle-\langle S_{i}{\sigma}_{i2}\rangle\right).
\label{md}
\end{eqnarray}  
It directly follows from Eq. (\ref{md}) that the decorating atoms oscillate symmetrically around their equilibrium lattice positions with  zero mean displacement $\langle\hat{\rho}_{i}\rangle=0$, because both nearest-neighbor pair correlation functions are identical $\langle S_{i}{\sigma}_{i1}\rangle = \langle S_{i}{\sigma}_{i2}\rangle$ and the mean displacement in a shifted coordinate system is null $\langle\hat{\rho}_{i}^{\prime}\rangle = 0$. It is thus necessary to compute the standard deviation for the displacement in order to shed light on oscillation of the decorating atoms around their equilibrium lattice positions
\begin{eqnarray}
d=\sqrt{\langle\hat{\rho}_{i}^{2}\rangle-\langle\hat{\rho}_{i}\rangle^{2}}=\sqrt{\langle\hat{\rho}_{i}^{2}\rangle}.
\label{rsm1}
\end{eqnarray}
The standard deviation for the displacement of the decorating atoms can be thus obtained from a square of the local canonical transformation (\ref{cct}) 
\begin{eqnarray}
\langle\hat{\rho}_{i}^{2}\rangle\!\!\!&=&\!\!\!\langle\hat{\rho}_{i}^{\prime 2}\rangle+\frac{A^2}{8K^2}\langle S_{i}^{2}\rangle-\frac{A^2}{2K^2}\langle S_{i}^{2}{\sigma}_{i1}{\sigma}_{i2}\rangle.
\label{rsm2}
\end{eqnarray}
The mean value for a square of the displacement of the decorating atoms from in a shifted coordinate system is given by 
\begin{eqnarray}
\langle \hat{\rho}_{i}^{\prime 2}\rangle=\frac{\hbar}{M\omega}\left(\left\langle\hat{b}_i^{+}\hat{b}_i^{-}\right\rangle+\frac{1}{2}\right)=\frac{\hbar\omega}{4K}\coth\left(\frac{\beta\hbar\omega}{2}\right),
\label{rsm3}
\end{eqnarray}
while the quadrupolar moment $\langle S_{i}^{2}\rangle$ and the three-site four-spin correlation function $\langle S_{i}^{2}{\sigma}_{i1}{\sigma}_{i2}\rangle$ can be calculated from the generalized Callen-Suzuki identity \cite{cal63,suz65,bal02} following the same approach as previously used for the calculation of uniform spontaneous magnetization of the decorating spins 
\begin{eqnarray}
\langle S_{i}^{2}\rangle\!\!\!&=&\!\!\!\frac{F_{1}+F_{2}}{2}+2\left(F_{1}-F_{2}\right)\varepsilon_{\rm IM}, \nonumber \\
\langle S_{i}^{2}{\sigma}_{i1}{\sigma}_{i2}\rangle\!\!\!&=&\!\!\!\frac{F_{1}-F_{2}}{8}+\frac{F_{1}+F_{2}}{2}\varepsilon_{\rm IM}.
\label{rsm4}
\end{eqnarray}
As before, the quantity $\varepsilon_{\rm IM}\equiv\langle{\sigma}_{i1}{\sigma}_{i2}\rangle$ denotes the nearest-neighbor pair correlation function of the corresponding spin-1/2 Ising model on undecorated lattice \cite{dom60} and the coefficients $F_1$ and $F_2$ were introduced in order to write the final formulas for the quadrupolar moment and the three-site four-spin correlation function in a more compact form
\begin{eqnarray}
F_{1}\!\!\!&=&\!\!\!\frac{\displaystyle\sum\limits_{n=-S}^{+S}n^2\exp\left(\beta Dn^{2}\right)\cosh\left(\beta Jn\right)}{\displaystyle\sum\limits_{n=-S}^{+S}\exp\left(\beta Dn^{2}\right)\cosh\left(\beta Jn\right)}, \nonumber \\
F_{2}\!\!\!&=&\!\!\!\frac{\displaystyle\sum\limits_{n=-S}^{+S}n^2\exp\left[\beta n^{2}(D+A^2/4K)\right]}{\displaystyle\sum\limits_{n=-S}^{+S}\exp\left[\beta n^{2}(D+A^2/4K)\right]}.
\label{rsm6}
\end{eqnarray}
After inserting the relations (\ref{rsm3})-(\ref{rsm6}) to the expression (\ref{rsm2}) one finally gets the final formula for the  standard deviation for the displacement of the decorating atoms 
\begin{eqnarray}
d\!\!\!&=&\!\!\!\sqrt{\frac{\hbar\omega}{4K}\coth\left(\frac{\beta\hbar\omega}{2}\right)+\frac{A^2}{8K^2}
\left(1-4\varepsilon_{IM}\right)F_{2}}. 
\label{sd}
\end{eqnarray} 

\section{Results and discussion}
\label{result}

Let us proceed to a discussion of the most interesting numerical results for the mixed spin-1/2 and spin-1 Ising model on a decorated square lattice being subject to lattice vibrations of the decorating atoms, which will bring insight into all important features of a more general model examined in the previous section for a quite general lattice geometry and the spin magnitude $S$. It is noteworthy that all results derived in Section \ref{model} reduce to the previously reported exact results for the mixed-spin Ising model on a rigid decorated planar lattice \cite{jas98,dak98} when considering a special limiting case of infinitely strong spring stiffness constant $K \to \infty$. In what follows, the equilibrium value of the nearest-neighbor bilinear interaction $|J|$ will be used as an energy unit when defining a relative strength of the uniaxial single-ion anisotropy $D/|J|$, the magnetoelastic coupling constant $A/|J|$, the bare elastic (spring stiffness) constant $K/|J|$, the characteristic frequency of normal mode oscillations $\hbar \omega/|J|$ and temperature $k_{\rm B} T/ |J|$ for both possible particular cases with either ferromagnetic $J>0$ or antiferromagnetic $J<0$ interaction. 

\begin{figure}
\begin{center}
\includegraphics[width=0.4\textwidth]{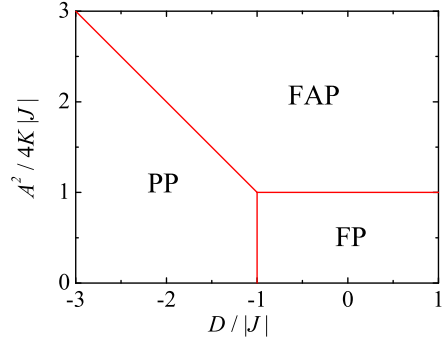}
\end{center}
\vspace{-1cm}
\caption{The ground-state phase diagram of the mixed spin-1/2 and spin-1 Ising model on a decorated square lattice in the 
$D/|J|-A^{2}/4K|J|$ plane. The notation for individual ground states is as follows: FP - ferromagnetic (ferrimagnetic) phase, 
PP - paramagnetic phase and FAP - frustrated antiferromagnetic phase.}
\label{fig2}       
\end{figure}

A survey of our theoretical results will be started with a detailed analysis of the ground-state phase diagram of the mixed spin-1/2 and spin-1 Ising model on a decorated square lattice, which is illustrated in Fig. \ref{fig2} in the $D/|J|-A^{2}/4K|J|$ plane. It should be pointed out that the displayed ground-state phase diagram is valid for both investigated particular cases with the ferromagnetic ($J>0$) as well as antiferromagnetic ($J<0$) pair interaction, whereas the relevant sign change is merely responsible for a change in a relative orientation of the decorating spins with respect to the nodal spins. It can be seen from Fig. \ref{fig2} that the nature of ground-state spin arrangement basically depends on whether a square of the magnetoelastic constant $A$ is smaller or greater than quadruple of a product between the bare elastic constant $K$ and the equilibrium exchange constant $|J|$. In the former case $A^2<4K|J|$, the spontaneously long-range ordered ferromagnetic or ferrimagnetic phase (FP) emerges for $D/|J|>-1$ depending on whether the equilibrium exchange constant is ferromagnetic $J>0$ or antiferromagnetic $J<0$, respectively. The disordered paramagnetic phase (PP) with nonmagnetic character of the decorating spins and paramagnetic character of the nodal spins appears in the remaining part of the parameter space $D/|J|<-1$. In the latter case $A^2>4K|J|$ the same disordered paramagnetic phase appears just at more negative values of the uniaxial single-ion anisotropy $D/|J|<-A^2/(4 K|J|)$, whereas the frustrated antiferromagnetic phase (FAP) with the disordered (paramagnetic) character of the decorating spins and the perfect antiferromagnetic long-range order of the nodal spins is being the respective ground state in the remaining part of the parameter space  $D/|J|>-A^2/(4 K|J|)$. The unusual frustrated antiferromagnetic phase displays qualitatively different behavior in comparison with the classical ferromagnetic and ferrimagnetic phases, so from now onward the relevant discussion will be split into two separate parts dealing with those two special cases.   

\subsection{Ferromagnetic (ferrimagnetic) phase}
\label{fero}

\begin{figure}
\begin{center}
\includegraphics[width=0.4\textwidth]{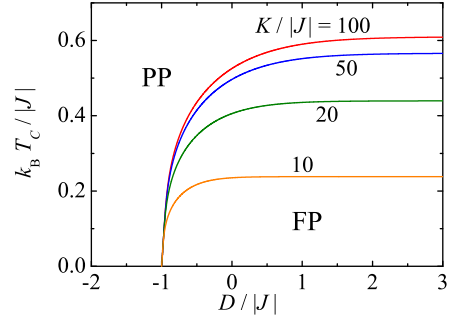}
\end{center}
\vspace{-1.1cm}
\caption{A critical temperature of the ferromagnetic (ferrimagnetic) phase as a function of the uniaxial single-ion anisotropy for the fixed value of the magnetoelastic coupling constant $A/|J|=5$ and several values ​​of the bare elastic constant $K/|J|$.} 
\label{fig3}       
\end{figure}

The finite-temperature phase diagram is shown in Fig.~\ref{fig3} in a form of critical temperature vs. uniaxial single-ion anisotropy plot for a set of the interaction parameters, which are consistent either with presence of the disordered paramagnetic phase for $D/|J|<-1$ or the spontaneously long-range ordered ferromagnetic (ferrimagnetic) phase for $D/|J|>-1$ and $J>0$ ($J<0$), respectively. It is noteworthy that the critical temperature is an even function of the equilibrium exchange constant $J$ as dictated by Eqs. (\ref{mp})-(\ref{v}) for the effective mapping parameter $\beta R_1$ and hence, the same critical line applies for the ferromagnetic ($J>0$) as well as ferrimagnetic ($J<0$) phase. It can be seen from Fig. \ref{fig3} that the critical temperature monotonically decreases by decreasing of the uniaxial single-ion anisotropy $D/|J|$ until it completely vanishes at the ground-state phase boundary $D/|J|=-1$ with the disordered paramagnetic phase. It can be also understood from Fig. \ref{fig3} that the critical temperature rises steadily upon strengthening of the bare elastic constant $K/|J|$, which means that the highest possible critical temperature is reached in the limit of a perfectly rigid lattice $K/|J| \to \infty$ \cite{jas98,dak98}. 

Next, let us take a closer look at temperature dependencies of the uniform spontaneous magnetization, which represents the relevant order parameter for the spontaneously long-range ordered ferromagnetic and ferrimagnetic phases. It should be remarked that the spontaneous sublattice magnetizations $m_A$ and $m_B$ of the nodal and decorating spins exhibit according to Eq. (\ref{mb}) the same thermal dependencies for both particular cases with the ferromagnetic ($J>0$) or antiferromagnetic ($J<0$) equilibrium exchange constant except a trivial change in the relative orientation of both sublattice magnetizations sign($m_A$) = sign($m_B$) for $J>0$ and sign($m_A$) = -sign($m_B$) for $J<0$, respectively. It is therefore quite instructive to examine first the relevant thermal variations of the spontaneous sublattice magnetizations $m_A$ and $m_B$ of the nodal and decorating spins in order to get a deeper insight into the total uniform spontaneous magnetization. 

\begin{figure}
\begin{center}
\includegraphics[width=0.4\textwidth]{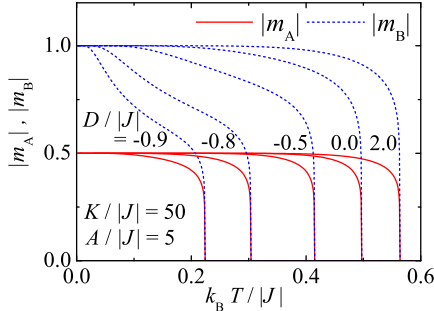}
\end{center}
\vspace{-0.9cm}
\caption{Thermal variations of the spontaneous sublattice magnetizations $|m_A|$ and $|m_B|$ of the nodal and decorating spins for the fixed value of the magnetoelastic coupling constant $A/|J|=5$, the bare elastic constant $K/|J|=50$ and several values ​​of the uniaxial single-ion anisotropy $D/|J|$.} 
\label{fig4}       
\end{figure}

It can be observed from Fig. \ref{fig4} that both spontaneous sublattice magnetizations $m_A$ and $m_B$ diminish likewise slightly below critical temperature regardless of a relative strength of the uniaxial single-ion anisotropy $D/|J|$, whereas the thermal behavior of the sublattice magnetizations $m_A$ and $m_B$ may display distinct features at low up to moderate temperatures. While the spontaneous sublattice magnetization $m_A$ is almost kept constant over a relatively wide temperature interval notwithstanding of the uniaxial single-ion anisotropy $D/|J|$, the spontaneous sublattice magnetization $m_B$ basically depends on the uniaxial single-ion anisotropy $D/|J|$. In fact, the more negative the uniaxial single-ion anisotropy is, the more rapid downturn of the sublattice magnetization $m_B$ can be detected at low up to moderate temperatures due to energetic favoring of a nonmagnetic state of the decorating spins.

\begin{figure}
\hspace*{-0.3cm}
\includegraphics[width=0.26\textwidth]{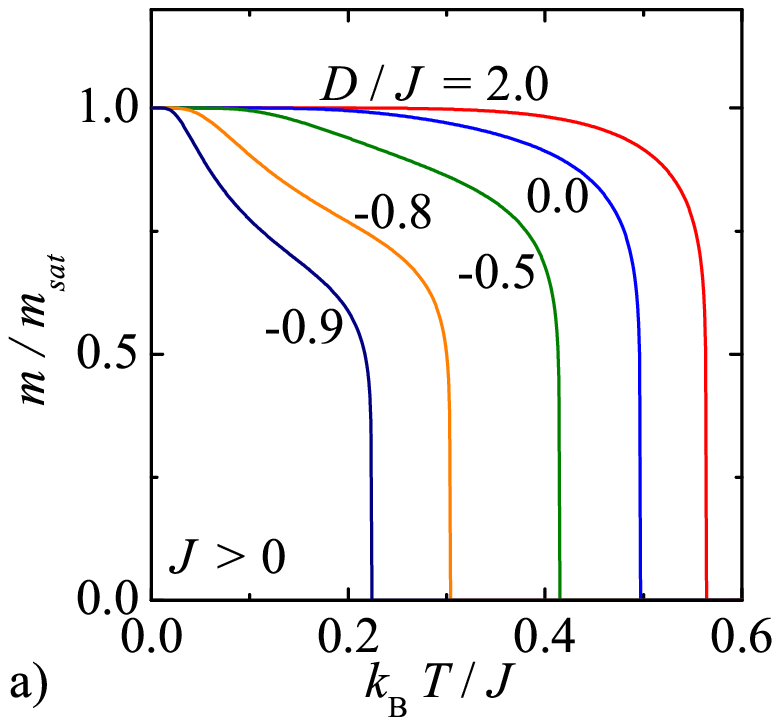}
\hspace*{-0.5cm}
\includegraphics[width=0.26\textwidth]{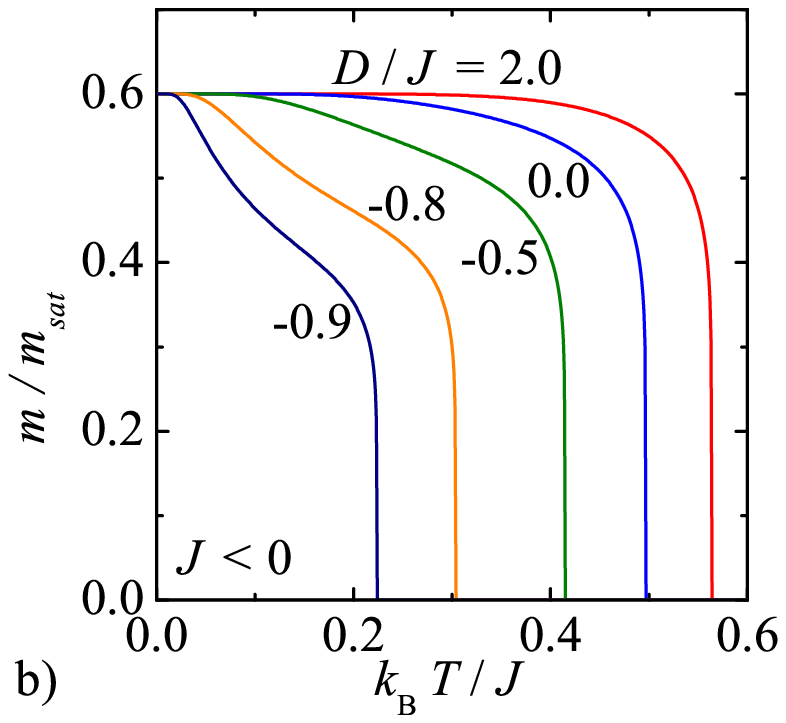}
\vspace{-0.9cm}
\caption{Thermal variations of the total magnetization normalized with respect to its saturation value for the magnetoelastic coupling constant $A/|J|=5$, the bare elastic constant $K/|J|=50$ and several values of the uniaxial single-ion anisotropy $D/|J|$ by considering the ferromagnetic $J>0$ [Fig. \ref{fig5}(a)] and antiferromagnetic $J<0$ [Fig. \ref{fig5}(b)] exchange constant.} 
\label{fig5}       
\end{figure}

The aforementioned trends in temperature dependencies of the spontaneous sublattice magnetizations $m_A$ and $m_B$ of the nodal and decorating spins have obvious impact upon thermal variations of the overall magnetization. The total spontaneous magnetization normalized  with respect to its saturation value is plotted in Fig. \ref{fig5} against temperature for several values of the uniaxial single-ion anisotropy by considering the ferromagnetic $J>0$ [Fig. \ref{fig5}(a)] and antiferromagnetic $J<0$ [Fig. \ref{fig5}(b)] exchange constant. The full saturation of the total magnetization observable in Fig. \ref{fig5}(a) in the asymptotic limit of zero temperature is in accordance with the perfect ferromagnetic long-range order of the nodal and decorating spins, while the partial saturation of the total magnetization observable in Fig. \ref{fig5}(b) agrees with the perfect ferrimagnetic long-range order of the nodal and decorating spins. In addition, it can be found that the negative values of the uniaxial single-ion anisotropy suppress not only critical temperature, but also the absolute value of the total magnetization at moderate temperatures due to the relevant thermal dependence of the sublattice magnetization $m_B$. 

\begin{figure}
\begin{center}
\includegraphics[width=0.24\textwidth]{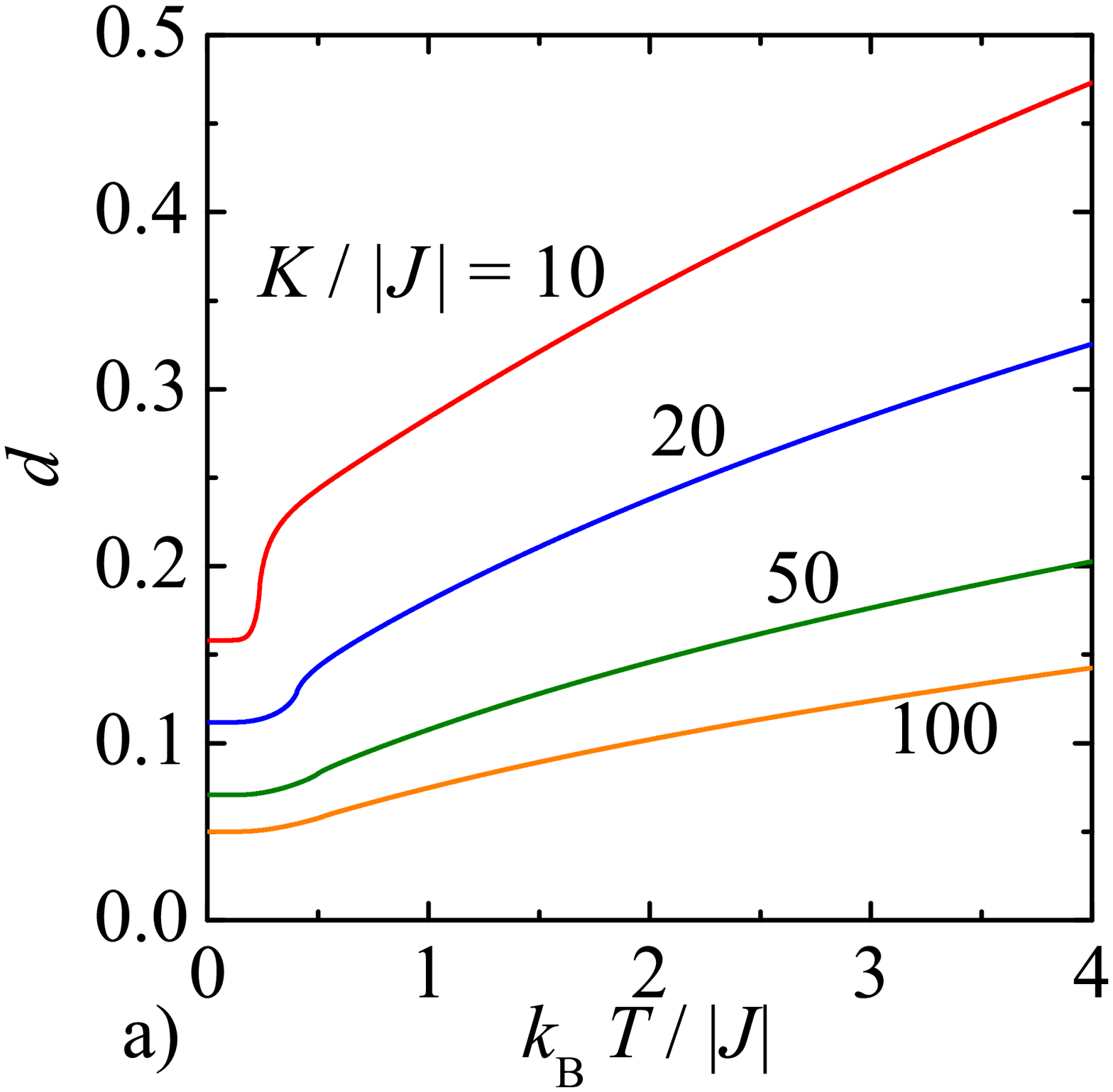}
\includegraphics[width=0.24\textwidth]{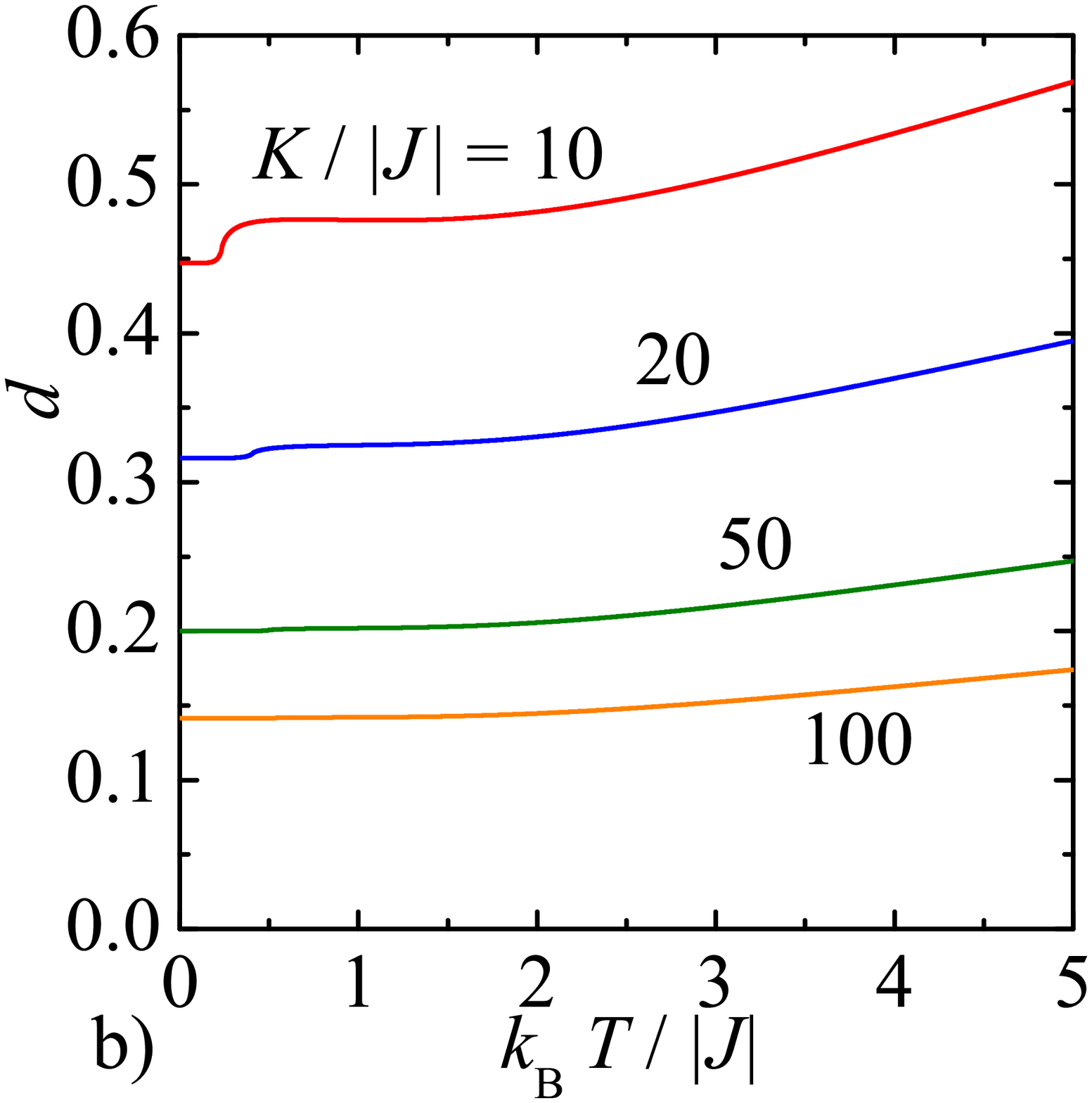}
\end{center}
\vspace{-0.7cm}
\caption{Temperature variations of the standard deviation for the displacement of the decorating atoms for the selected value of the magnetoelastic constant $A/|J|=5$, the  uniaxial single-ion anisotropy $D/|J|=0$ and two angular frequencies of the normal-mode oscillations: 
a) $\hbar\omega/|J|=1$, b) $\hbar\omega/|J|=8$.} 
\label{fig6}       
\end{figure}

To verify a range of applicability of the harmonic approximation the standard deviation for the displacement of the decorating atoms is plotted in Fig. \ref{fig6} against temperature for a few selected values of the bare elastic constant $K/|J|$ and two angular frequencies of normal-mode oscillations. Notice that the standard deviation for the displacement depends according to Eq. (\ref{sd}) on the nearest-neighbor pair correlation function $\varepsilon_{\rm IM}\equiv\langle{\sigma}_{i1}{\sigma}_{i2}\rangle$, which is also through the effective mapping parameter $\beta R_1$ (\ref{mp})-(\ref{v}) an even function of the equilibrium exchange constant $J$. This result is taken to mean that the displayed thermal dependencies of the standard deviation hold both for the ferromagnetic ($J>0$) as well as antiferromagnetic ($J<0$) exchange constant. It is quite clear from Fig. \ref{fig6} that the standard deviation for the displacement of the decorating atoms remains reasonably small for sufficiently stiff lattices with the bare elastic constant $K/|J| \gtrsim 50$ within 
the whole temperature range, which is of particular  interest with regard to a nontrivial spontaneous long-range magnetic order.

\begin{figure}
\begin{center}
\includegraphics[width=0.24\textwidth]{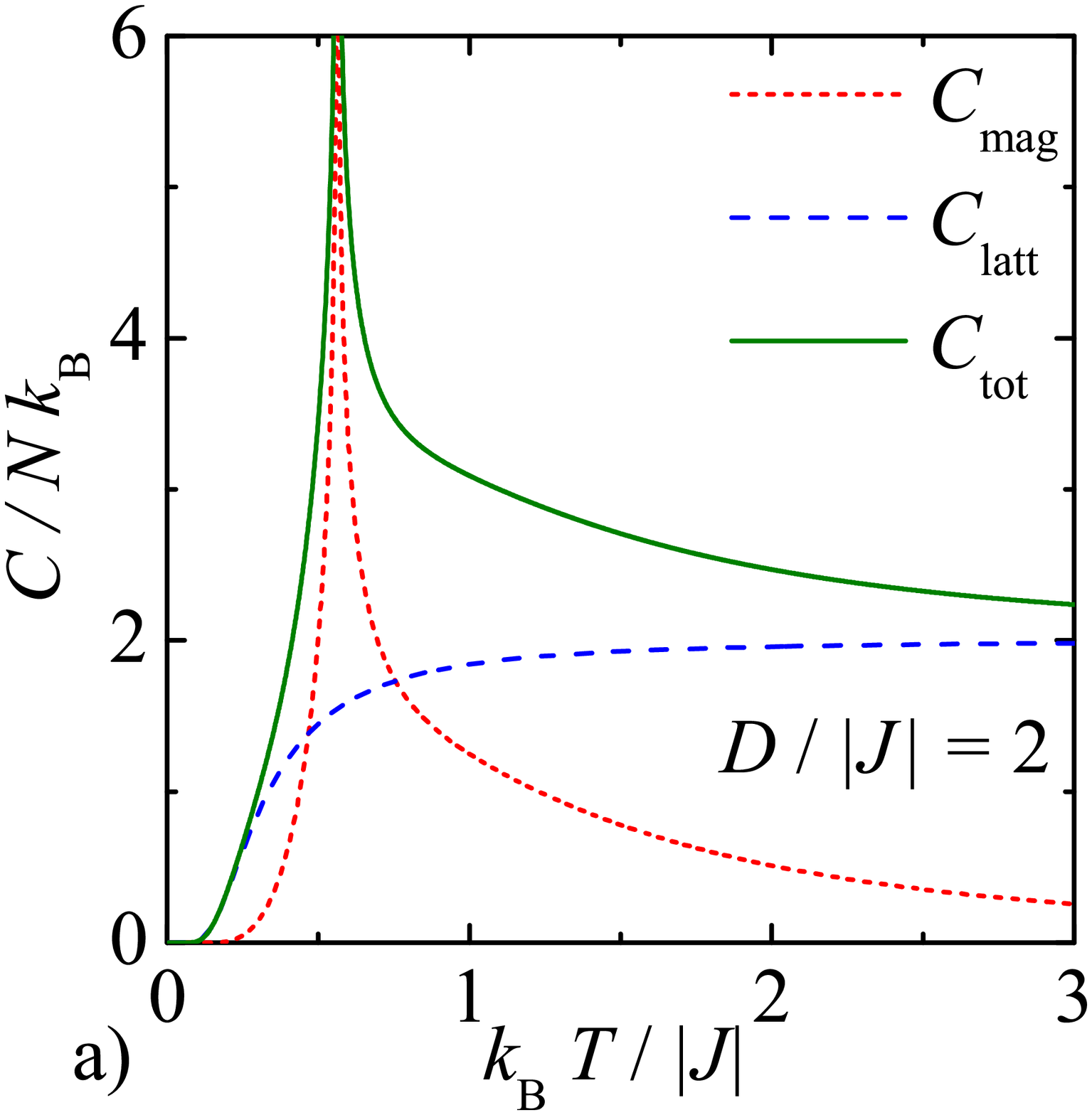}
\includegraphics[width=0.24\textwidth]{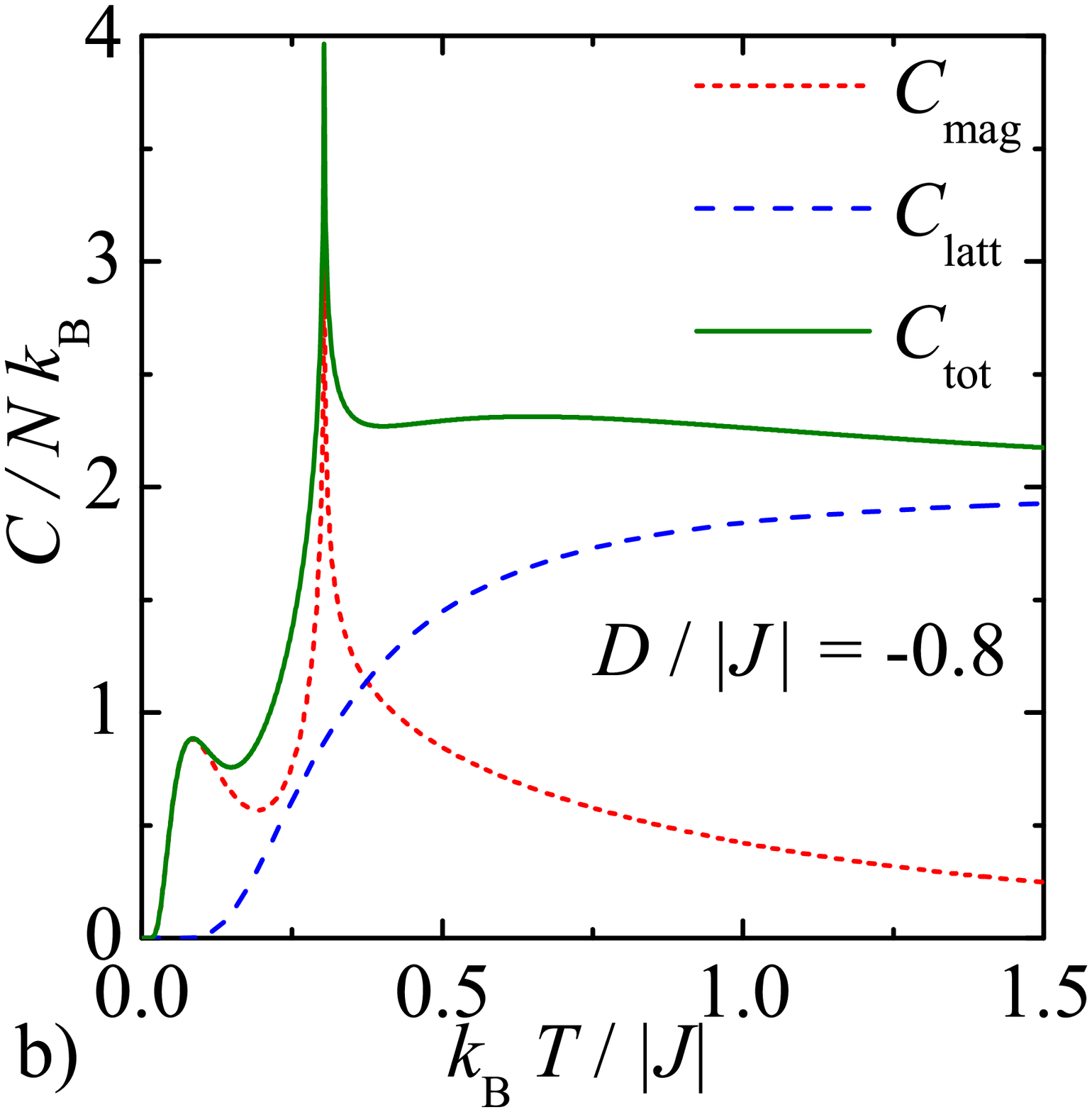}
\includegraphics[width=0.24\textwidth]{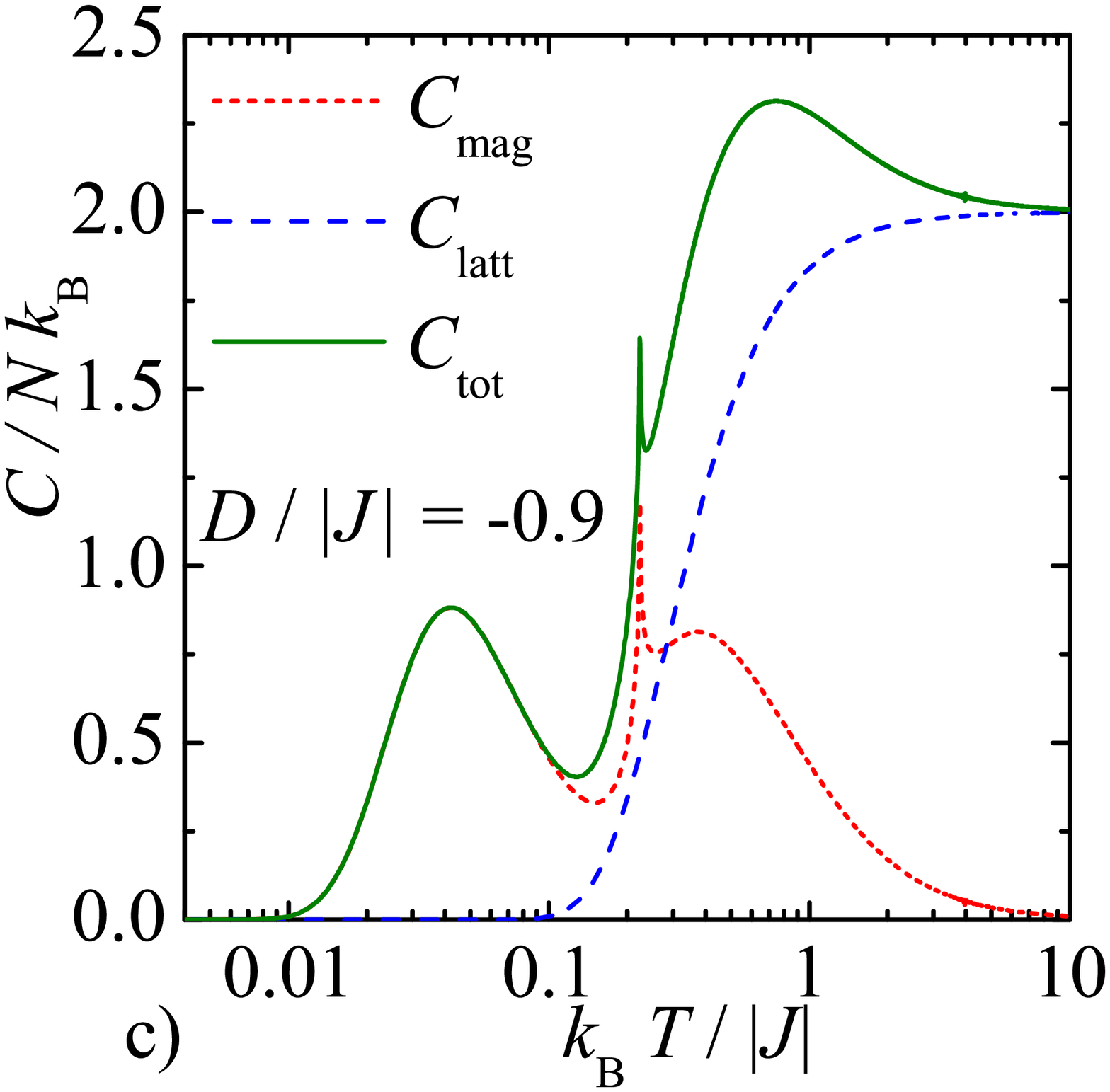}
\includegraphics[width=0.24\textwidth]{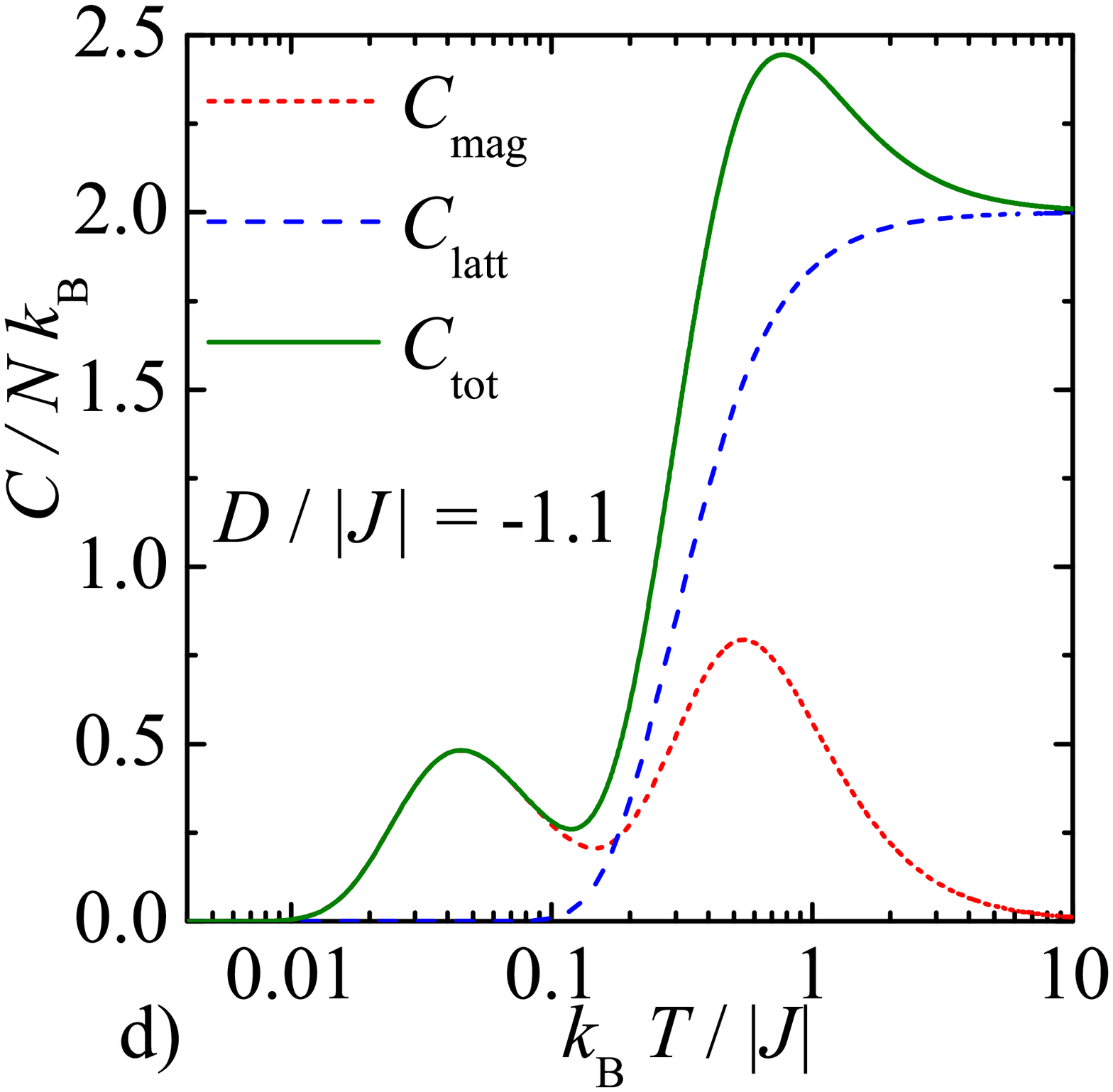}
\end{center}
\vspace{-0.7cm}
\caption{Typical temperature dependencies of the total, magnetic and lattice parts of the specific heat for the selected value of the magnetoelastic coupling constant $A/|J|=5$, the angular frequency of normal-mode oscillations $\hbar\omega/|J|=1$, the bar elastic constant 
$K/|J|=50$ and several values of the uniaxial single-ion anisotropy: a) $D/|J|=2.0$, b) $D/|J|=-0.8$, c) $D/|J|=-0.9$, d) $D/|J|$=-1.1.} 
\label{fig7}       
\end{figure}

\begin{figure}
\begin{center}
\includegraphics[width=0.24\textwidth]{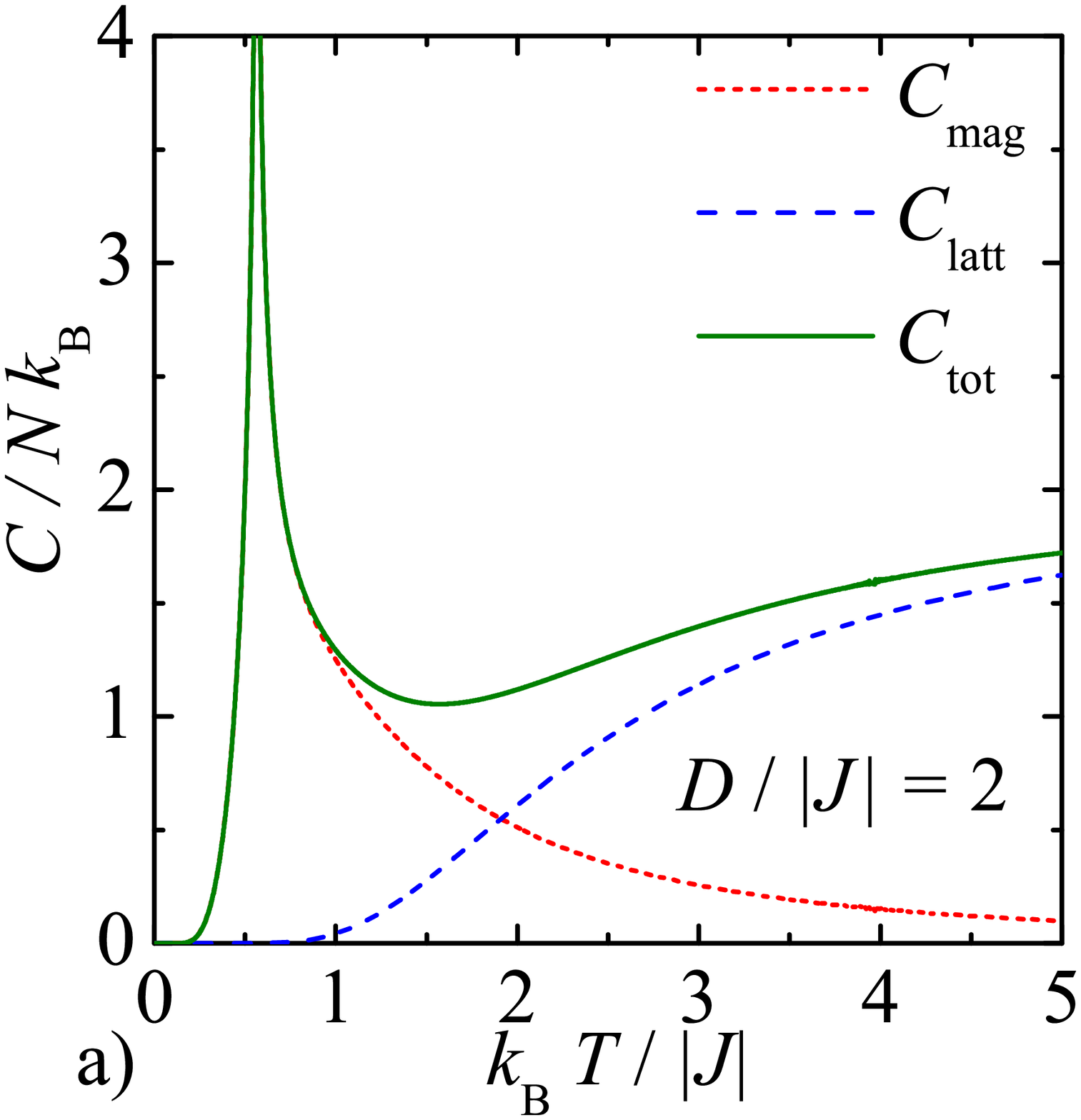}
\includegraphics[width=0.24\textwidth]{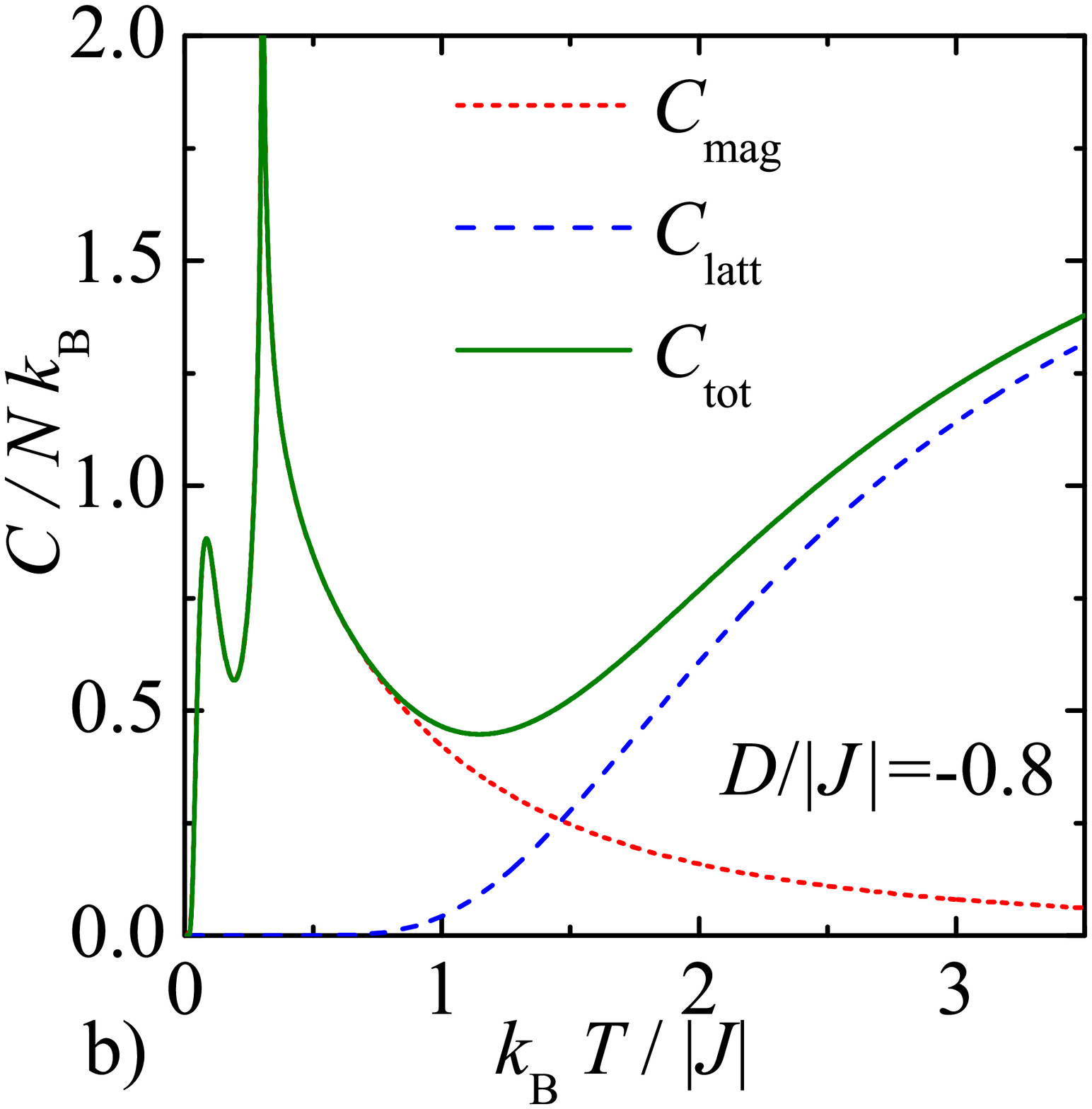}
\includegraphics[width=0.24\textwidth]{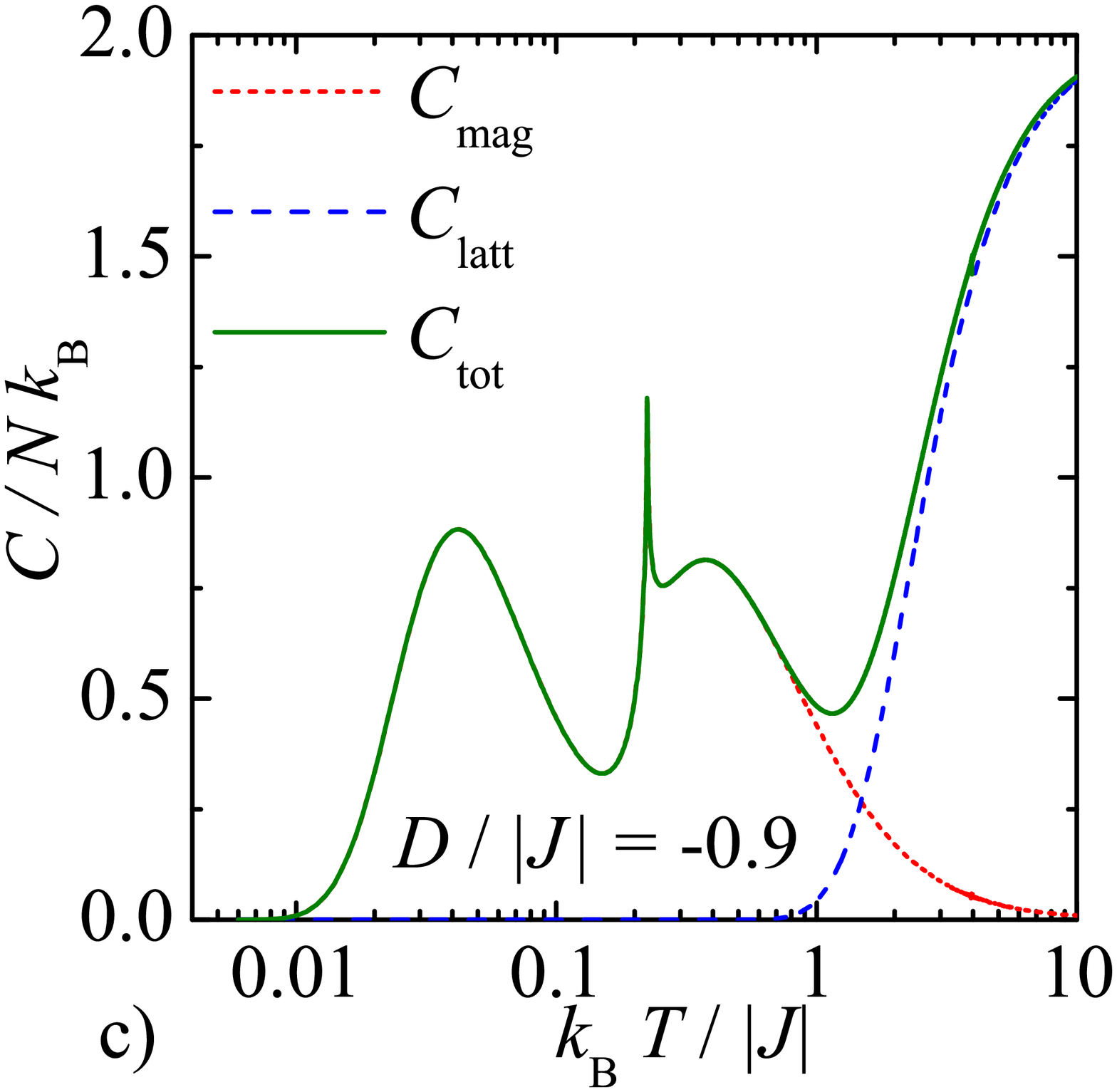}
\includegraphics[width=0.24\textwidth]{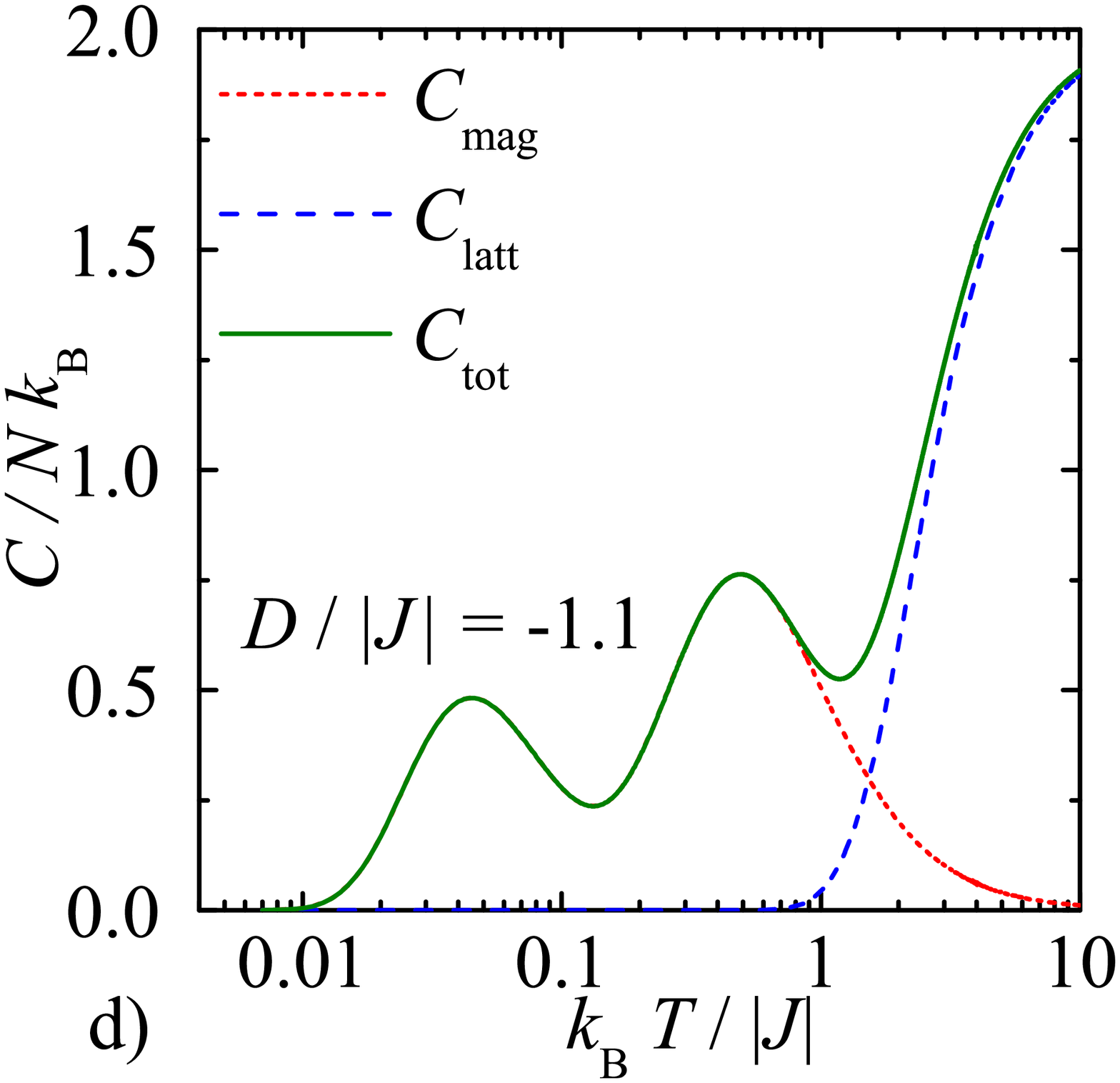}
\end{center}
\vspace{-0.7cm}
\caption{Typical temperature dependencies of the total, magnetic and lattice parts of the specific heat for the selected value of the magnetoelastic coupling constant $A/|J|=5$, the angular frequency of normal-mode oscillations $\hbar\omega/|J|=8$, the bar elastic constant 
$K/|J|=50$ and several values of the uniaxial single-ion anisotropy: a) $D/|J|=2.0$, b) $D/|J|=-0.8$, c) $D/|J|=-0.9$, d) $D/|J|$=-1.1.} 
\label{fig8}       
\end{figure}

Last but not least, let us investigate in detail typical temperature dependencies of the specific heat, which are plotted in Figs. \ref{fig7} and \ref{fig8} together with its magnetoelastic and pure lattice contributions for several values of the uniaxial single-ion anisotropy. For positive or weakly negative uniaxial single-ion anisotropies the overall specific heat exhibits at a critical temperature a standard logarithmic divergence stemming from its magnetic contribution, whereas the high-temperature tail of $\lambda$-anomaly is either shifted upwards [Figs. \ref{fig7}(a)] or is superimposed on ascending part of the lattice contribution [Figs. \ref{fig8}(a)] depending on the angular frequency of normal-mode oscillations. The overall specific heat displays similar temperature dependencies also at moderate negative values of the uniaxial single-ion anisotropy except that an additional round maximum is formed in a low-temperature tail of the specific heat [see Figs. \ref{fig7}(b),(c) and \ref{fig8}(b),(c)]. The low-temperature round maximum is apparently of a magnetic origin and it can be explained as the Schottky-type maximum arising from two closely spaced energy levels of the decorating spins. The total specific heat is free of any divergence at more negative values of the uniaxial single-ion anisotropy as exemplified in Figs. \ref{fig7}(d) and \ref{fig8}(d). Two round maxima of the magnetic origin emerge either before the ascending part of the lattice contribution [Fig. \ref{fig7}(d)] or the latter round maximum may be superimposed on the top of the lattice contribution [Fig. \ref{fig8}(d)].

\subsection{Frustrated antiferromagnetic phase}
\label{antifero}

\begin{figure}
\begin{center}
\includegraphics[width=0.4\textwidth]{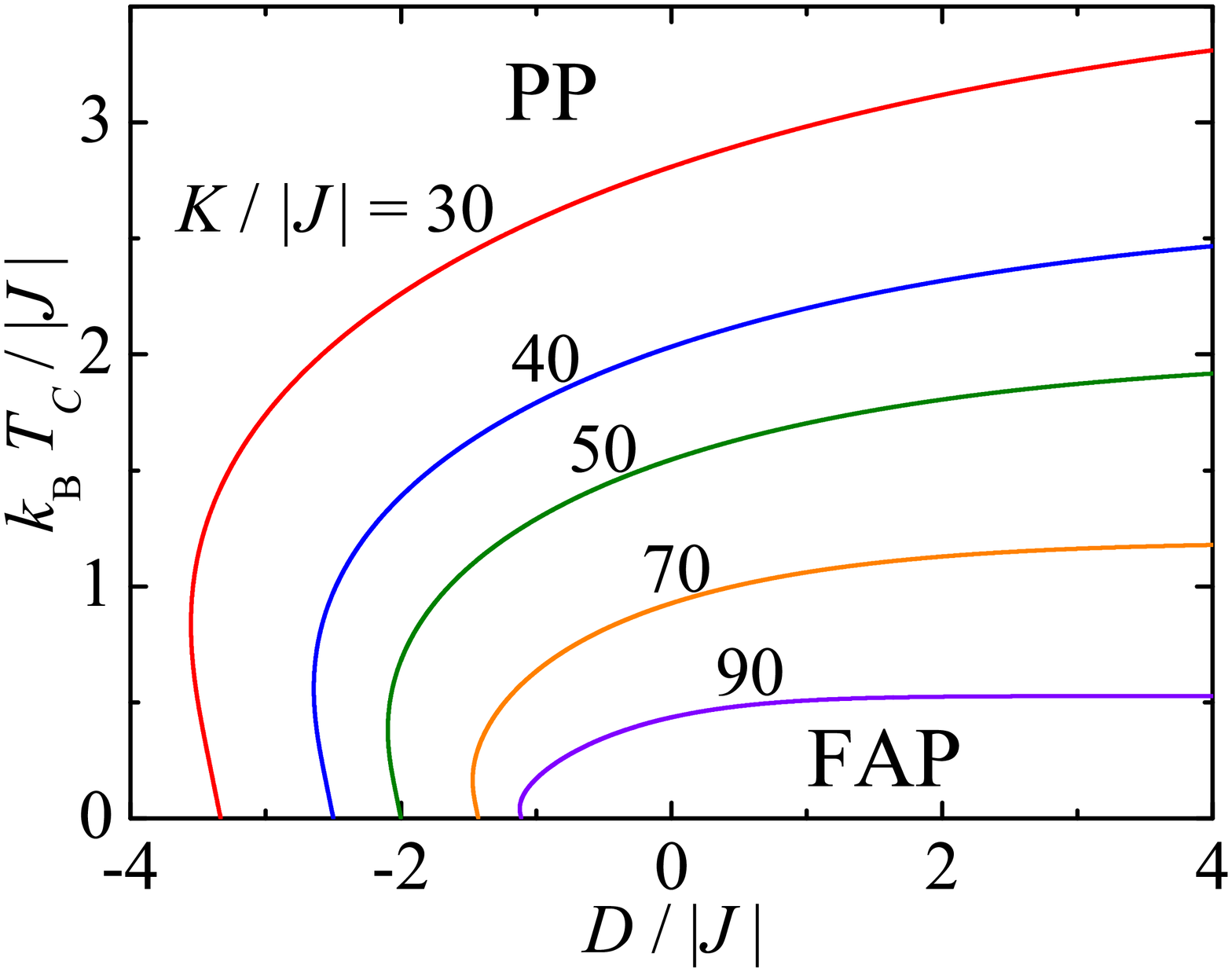}
\end{center}
\vspace{-0.9cm}
\caption{A critical temperature of the frustrated antiferromagnetic phase as a function of the uniaxial single-ion anisotropy for the fixed value of the magnetoelastic coupling constant $A/|J|=20$ and several values ​​of the bare elastic constant $K/|J|$.} 
\label{fig9}       
\end{figure}

In this section we will explore the main features of the investigated spin system when driven by a suitable choice of the interaction parameters towards the frustrated antiferromagnetic phase. Fig.~\ref{fig9} displays the critical temperature against the uniaxial single-ion anisotropy for a set of the interaction parameters, which either lead to the disordered paramagnetic phase for $D/|J|<-A^2/(4K|J|)$ or the frustrated antiferromagnetic phase for $D/|J|>-A^2/(4K|J|)$. It is worthwhile to recall that the critical temperature is independent of a sign of the equilibrium exchange constant $J$, which may be chosen arbitrarily without any effect upon the character of the frustrated antiferromagnetic phase. In contrast to the previous case, the critical temperature of the frustrated antiferromagnetic phase does not monotonically fall down upon decreasing of the uniaxial single-ion anisotropy $D/|J|$, but it instead shows a double reentrant behavior in a vicinity of the ground-state phase boundary $D/|J|=-A^2/(K|J|)$ with the disordered paramagnetic phase. The double reentrant phase transitions can be attributed to a relatively high entropy the frustrated antiferromagnetic phase with a paramagnetic (frustrated) character of the decorating spins and a perfect antiferromagnetic long-range order of the nodal spins. Another interesting observation is that the reentrant region as well as the critical temperature generally shrink upon strengthening of the lattice stiffness $K/|J|$. 

\begin{figure}
\begin{center}
\includegraphics[width=0.4\textwidth]{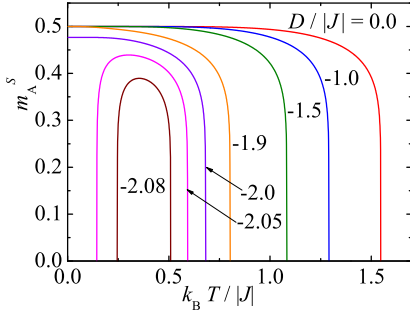}
\end{center}
\vspace{-0.9cm}
\caption{The spontaneous staggered magnetization of the nodal spins as a function of temperature for the magnetoelastic coupling constant 
$A/|J|=20$, the bare elastic constant $K/|J|=50$ and several values ​​of the uniaxial single-ion anisotropy.} 
\label{fig10}       
\end{figure}

A peculiar nature of the frustrated antiferromagnetic phase consists in a coexistence of a partial disorder of the decorating spins randomly occupying one of two magnetic states with a perfect antiferromagnetic order of the nodal spins. The spontaneous staggered magnetization of the nodal spins $m_A^s$ thus represents a relevant order parameter of the frustrated antiferromagnetic phase, whose typical temperature plots are depicted in Fig. \ref{fig10}. If the uniaxial single-ion anisotropy $D/|J|>-A^2/(4K|J|)$ enforces the frustrated antiferromagnetic ground state, then, the spontaneous staggered magnetization of the nodal spins starts from its maximum possible value and is subsequently reduced upon increasing of temperature until it completely vanishes at a critical temperature. This thermal behavior is strongly reminiscent of that reported previously for the spontaneous uniform magnetization of the nodal spins within the ferromagnetic (ferrimagnetic) phase. However, there also may appear remarkable temperature variations of the staggered magnetization of the nodal spins when the uniaxial single-ion anisotropy $D/|J| \lesssim -A^2/(4K|J|)$ prefers the disordered paramagnetic ground state but is still sufficiently close to the ground-state phase boundary with the frustrated antiferromagnetic phase. Under this condition, the staggered magnetization of the nodal spins shows a remarkable double reentrant behavior when it only appears at a lower critical temperature and disappears at an upper critical temperature (see the dependencies for $D/|J| = -2.05$ and $-2.08$ in Fig. \ref{fig10}). 

\begin{figure}
\begin{center}
\includegraphics[width=0.24\textwidth]{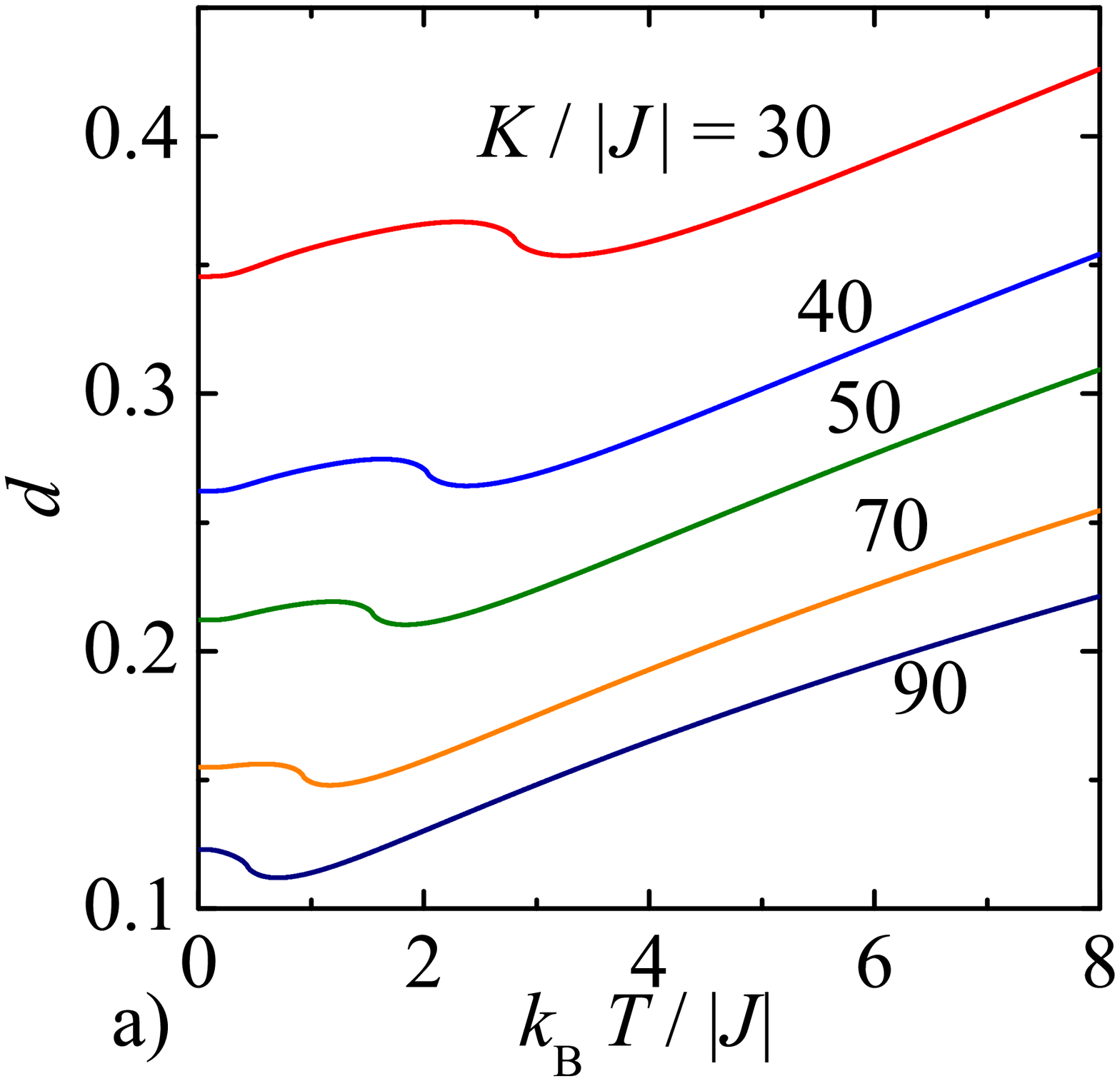}
\includegraphics[width=0.24\textwidth]{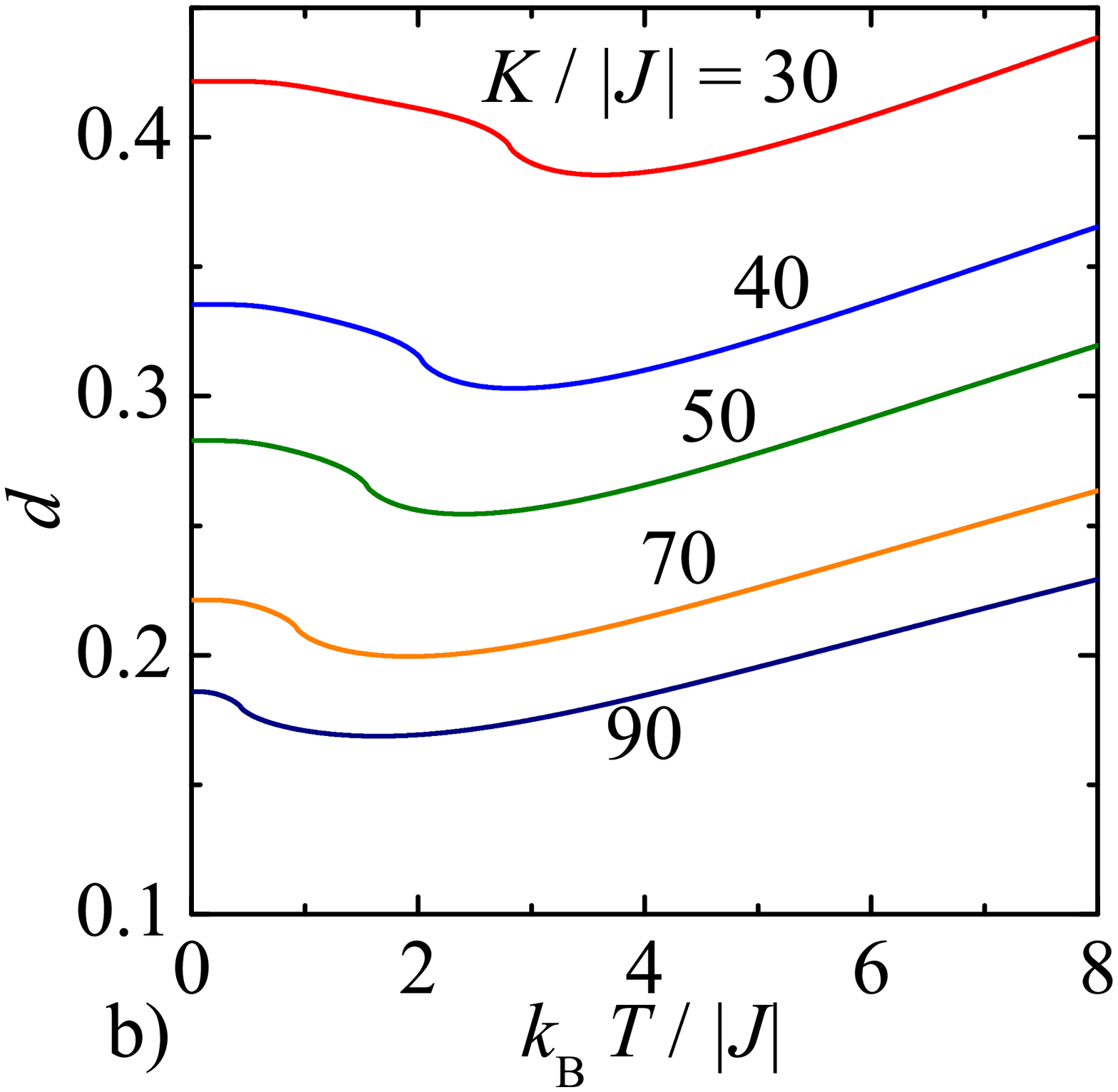}
\end{center}
\vspace{-0.7cm}
\caption{Temperature variations of the standard deviation for the displacement of the decorating atoms for the selected value of the magnetoelastic constant $A/|J|=20$, the  uniaxial single-ion anisotropy $D/|J|=0$ and two angular frequencies of the normal-mode oscillations: 
a) $\hbar\omega/|J|=1$, b) $\hbar\omega/|J|=8$.} 
\label{fig11}       
\end{figure}

Temperature variations of the standard deviation for the displacement of the decorating atoms shown in Fig. \ref{fig11} allow us to estimate  a validity of the harmonic approximation. Recall that the displayed thermal dependencies of the standard deviation are valid both for the ferromagnetic ($J>0$) as well as antiferromagnetic ($J<0$) exchange constant. It directly follows from Fig. \ref{fig11} that the standard deviation for the displacement of the decorating atoms is small enough within the adequate temperature range for sufficiently stiff lattices with the bare elastic constant $K/|J| \gtrsim 50$. It is interesting to notice, however, that the standard deviation for the displacement of the decorating atoms exhibits a striking nonmonotonous temperature dependence with a weak energy-type singularity at a relevant critical temperature. 

\begin{figure}
\begin{center}
\includegraphics[width=0.24\textwidth]{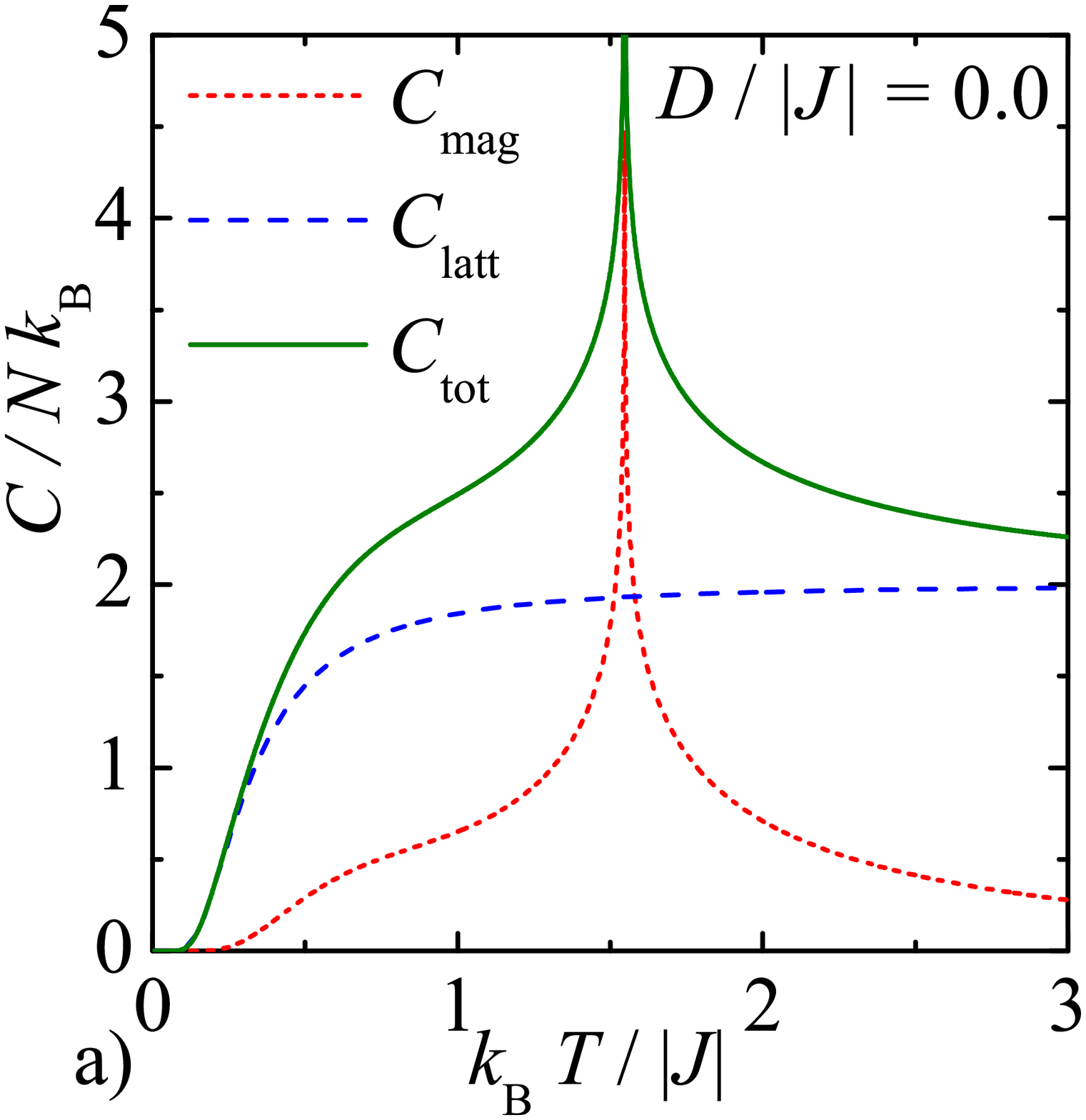}
\includegraphics[width=0.24\textwidth]{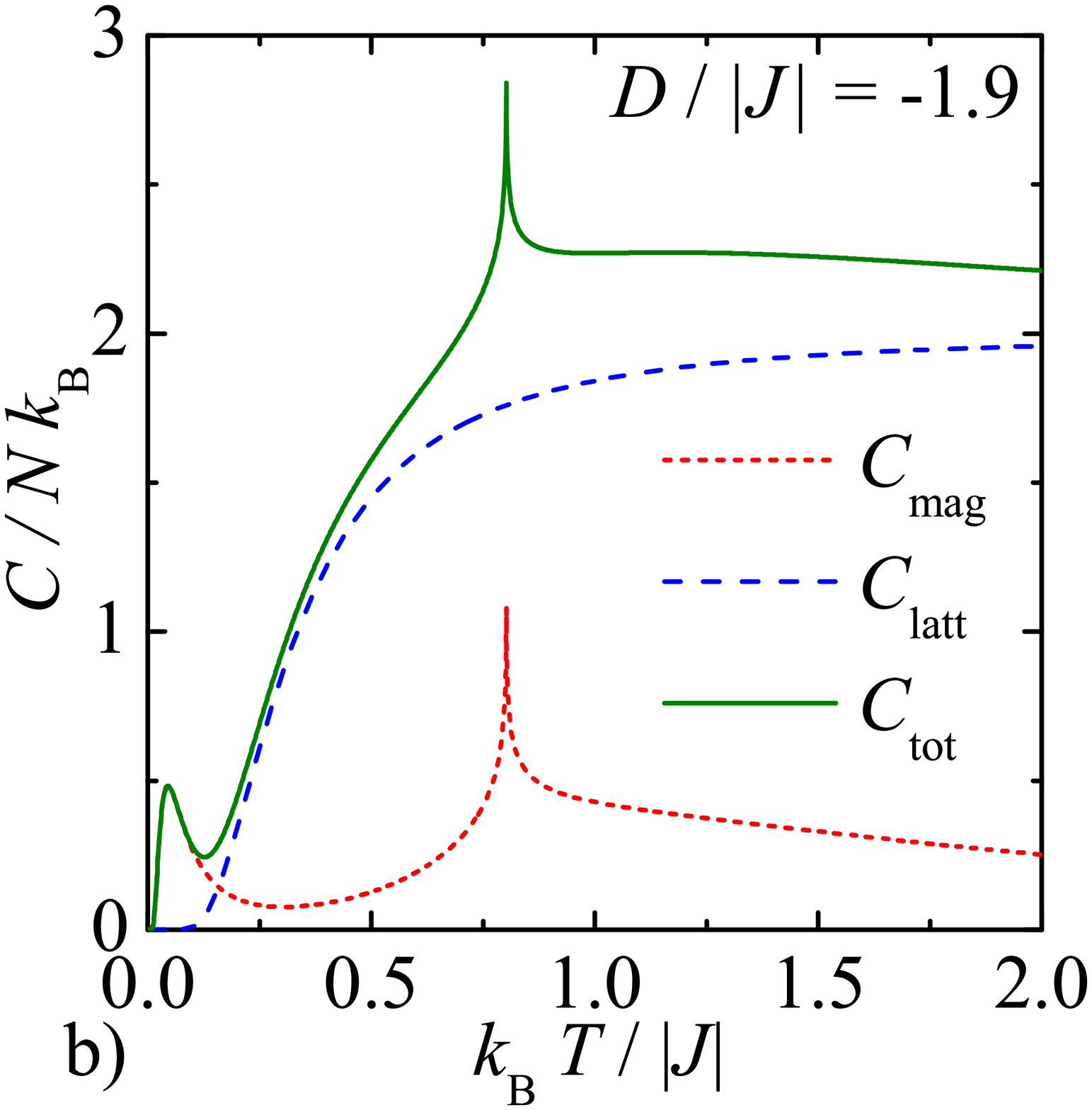}
\includegraphics[width=0.24\textwidth]{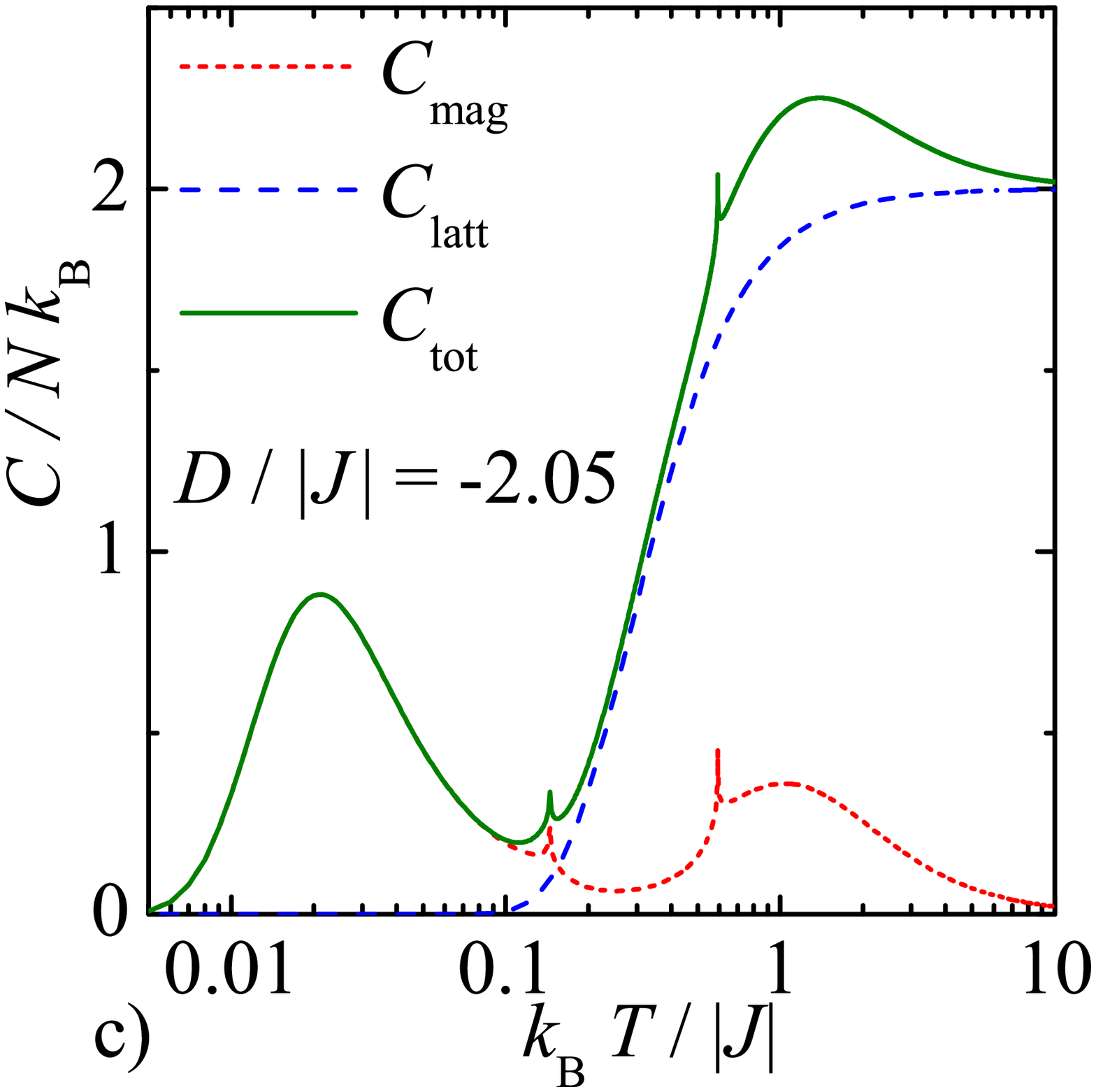}
\includegraphics[width=0.24\textwidth]{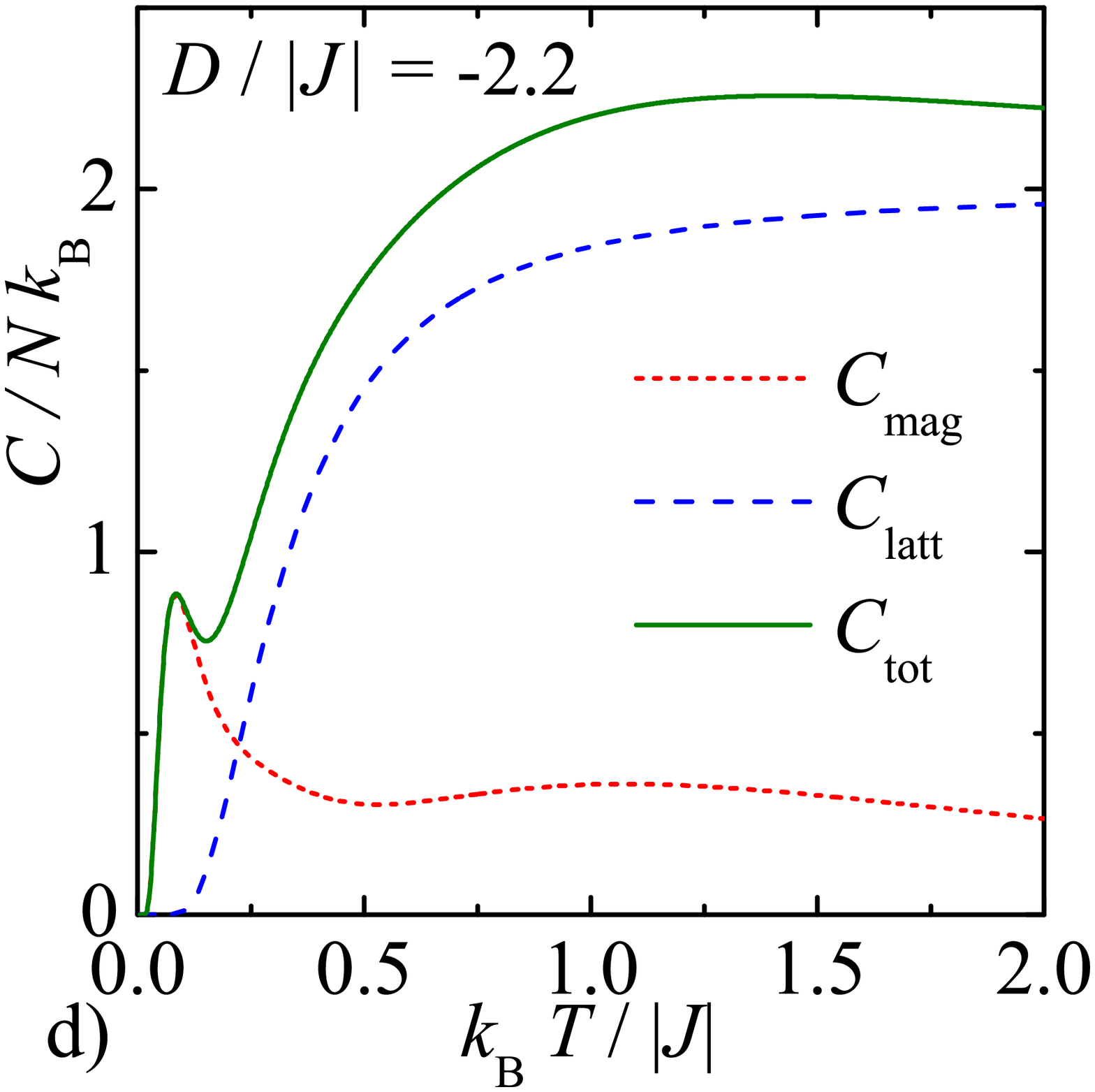}
\end{center}
\vspace{-0.7cm}
\caption{Typical temperature dependencies of the total, magnetic and lattice parts of the specific heat for the selected value of the magnetoelastic coupling constant $A/|J|=20$, the angular frequency of normal-mode oscillations $\hbar\omega/|J|=1$, the bare elastic constant 
$K/|J|=50$ and several values of the uniaxial single-ion anisotropy: a) $D/|J|=0.0$, b) $D/|J|=-1.9$, c) $D/|J|=-2.05$, d) $D/|J|$=-2.2.} 
\label{fig12}       
\end{figure}

\begin{figure}
\begin{center}
\includegraphics[width=0.24\textwidth]{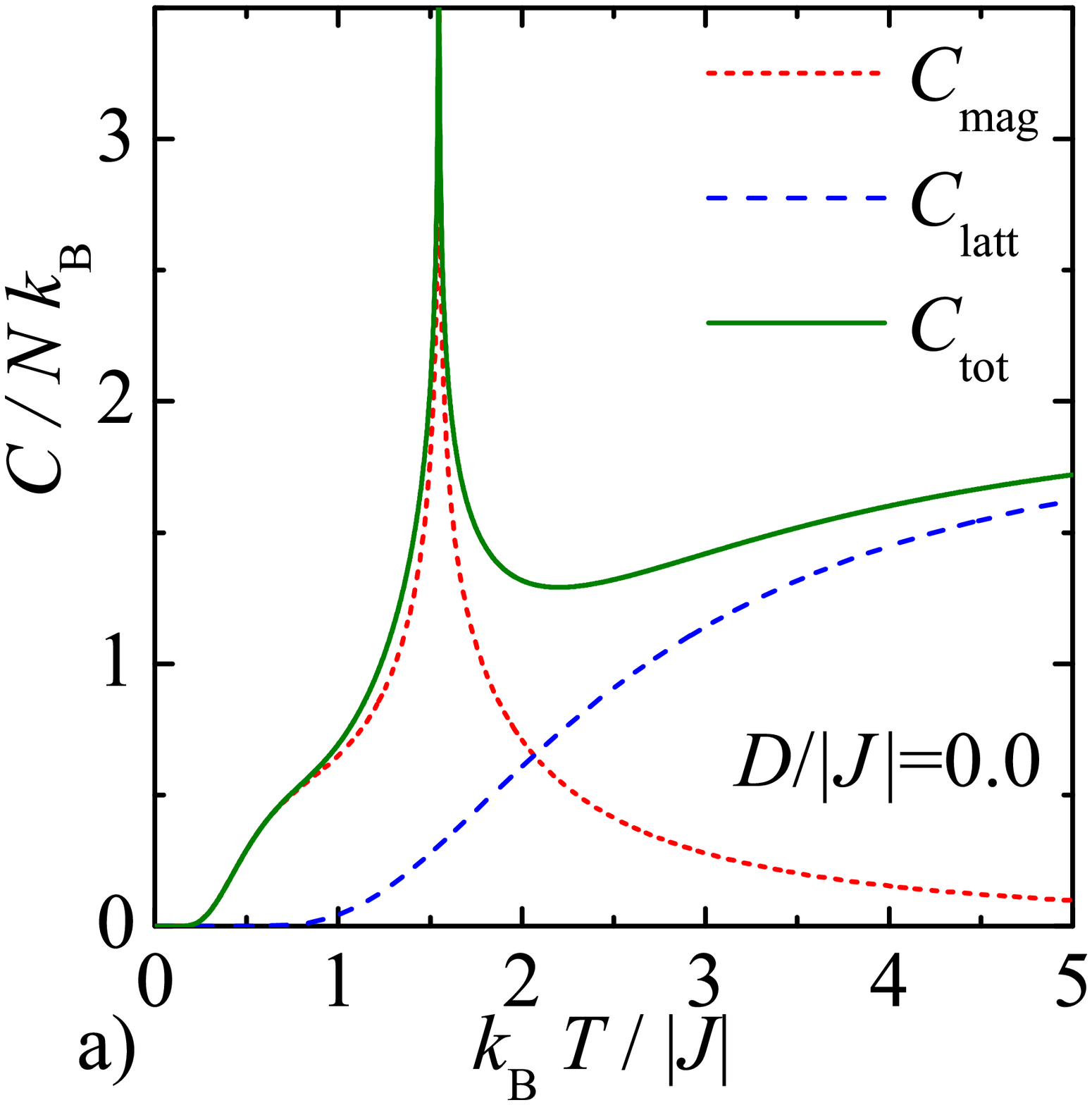}
\includegraphics[width=0.24\textwidth]{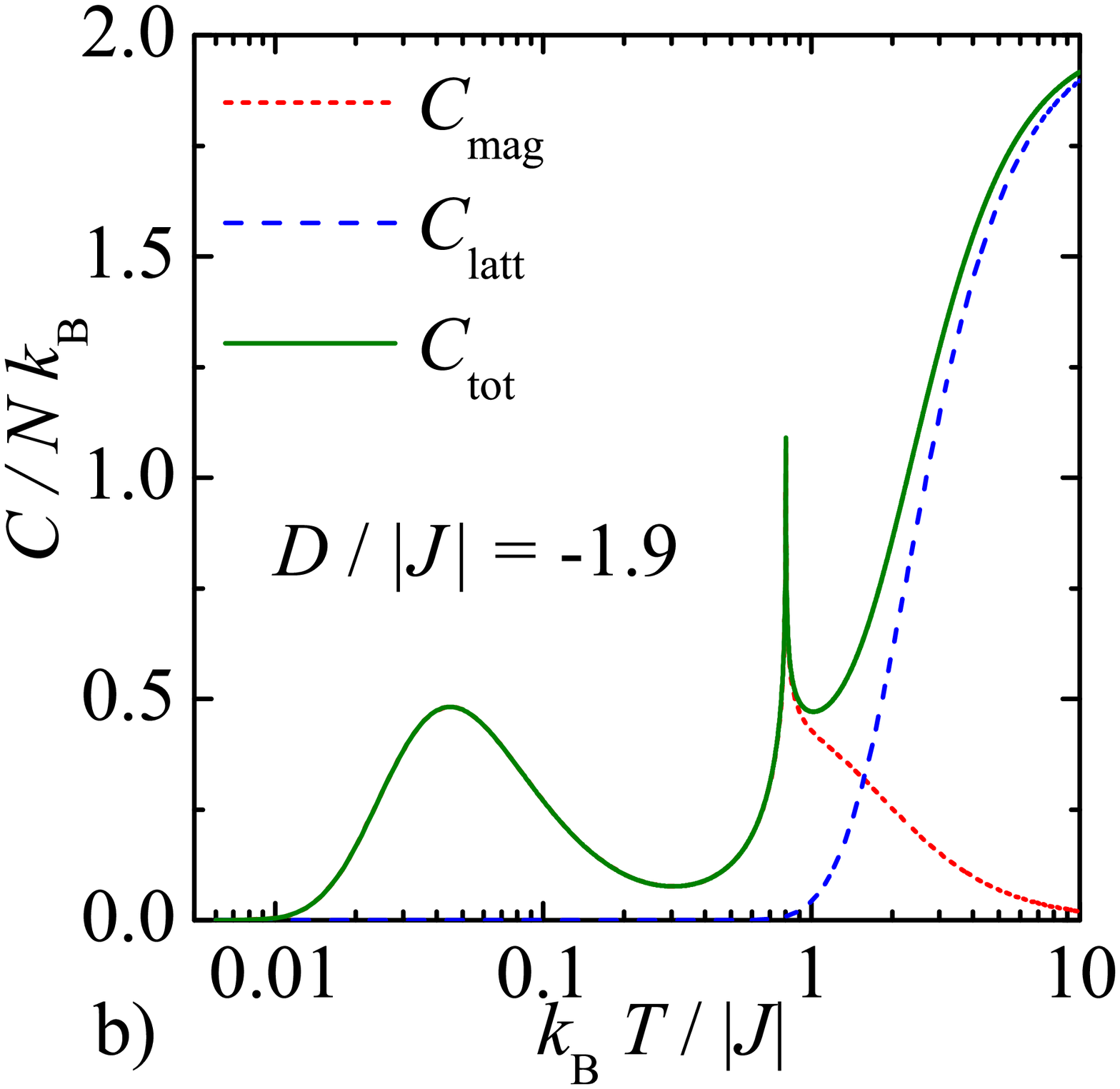}
\includegraphics[width=0.24\textwidth]{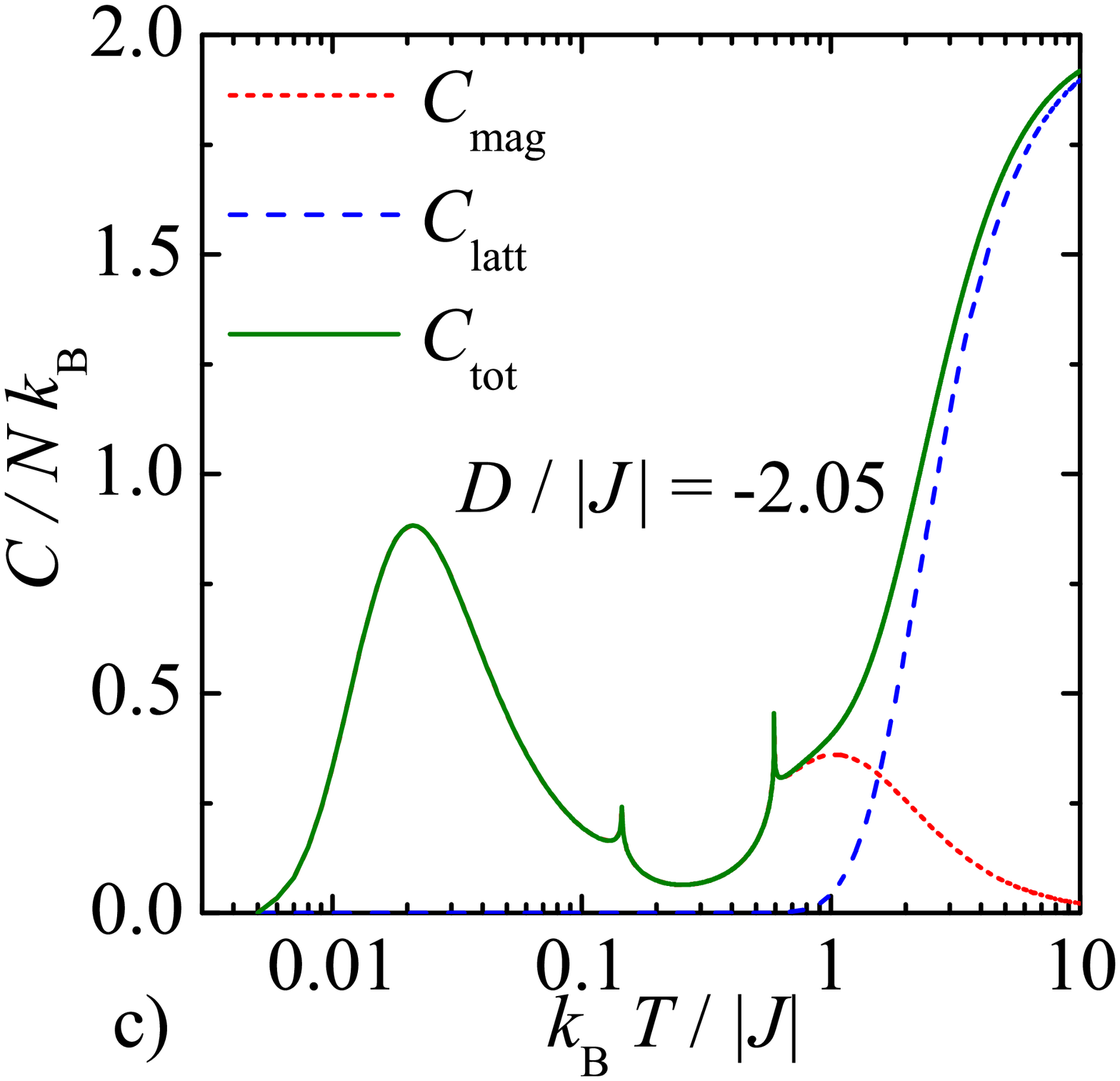}
\includegraphics[width=0.24\textwidth]{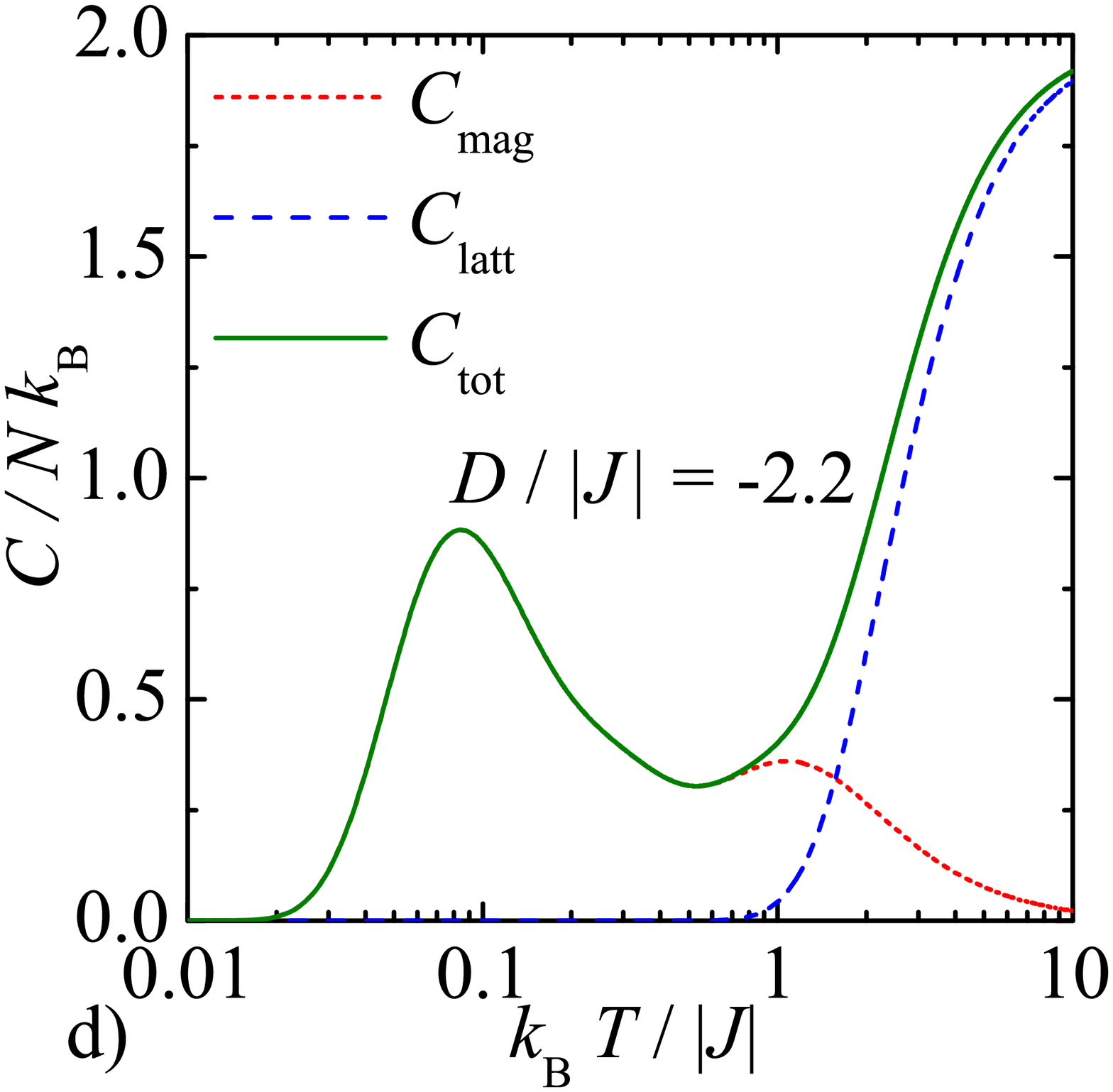}
\end{center}
\vspace{-0.7cm}
\caption{Typical temperature dependencies of the total, magnetic and lattice parts of the specific heat for the selected value of the magnetoelastic coupling constant $A/|J|=20$, the angular frequency of normal-mode oscillations $\hbar\omega/|J|=8$, the bare elastic constant 
$K/|J|=50$ and several values of the uniaxial single-ion anisotropy: a) $D/|J|=0.0$, b) $D/|J|=-1.9$, c) $D/|J|=-2.05$, d) $D/|J|$=-2.2.} 
\label{fig13}       
\end{figure}

Let us conclude our survey of the magnetic behavior of the frustrated antiferromagnetic phase by exploring typical temperature dependencies of the total specific heat, which is displayed in Figs. \ref{fig12} and \ref{fig13} along with its magnetoelastic and pure lattice contributions for several values of the uniaxial single-ion anisotropy. It can be seen from Figs. \ref{fig12}(a) and \ref{fig13}(a) that the logarithmic divergence descended from a magnetic part of the specific heat may appear either on a saturation part of the lattice contribution [Fig. \ref{fig12}(a)] or superimposed on an ascending part of the lattice contribution [Fig. \ref{fig13}(a)] depending on a relative size of the angular frequency of the normal-mode oscillations. This statement holds true for positive, zero or weakly negative values of the uniaxial single-ion anisotropy. The most peculiar temperature dependencies of the specific heat can be however detected if the uniaxial single-ion anisotropy drives the investigated spin system sufficiently close to the ground-state phase boundary $D/|J| = -A^2/(4K|J|)$ between the frustrated antiferromagnetic phase and the disordered paramagnetic phase. The total specific heat then exhibits an additional round maximum at low enough temperatures, which can be repeatedly interpreted as the Schottky-type maximum of a magnetic origin arising from two closely spaced energy levels of the decorating spins [see Figs. \ref{fig12}(b),(c) and \ref{fig13}(b),(c)]. In addition, two logarithmic divergences of the magnetic specific heat in Figs. \ref{fig12}(c) and \ref{fig13}(c) provide a further evidence of the double reentrant critical behavior emergent for the uniaxial single-ion anisotropies $D/|J| \lesssim -A^2/(4K|J|)$ chosen slightly below the ground-state phase boundary with the frustrated antiferromagnetic phase. At more negative values of the uniaxial single-ion anisotropy both logarithmic divergences of the specific heat naturally disappear due to a lack of spontaneous long-range order suppressed by the disordered paramagnetic phase [see Figs. \ref{fig12}(d) and \ref{fig13}(d)]. 

\section{Concluding remarks}
\label{conclusion}

In the present paper we have examined magnetoelastic properties of the mixed-spin Ising model on decorated planar lattices when accounting for lattice vibrations of the decorating atoms within the canonical coordinate transformation, the decoration-iteration transformation, and the harmonic approximation. It has been found that the magnetoelastic coupling gives rise to an effective single-ion anisotropy and three-site four-spin interaction, which may compete with the equilibrium bilinear exchange constant and may enforce the anomalous frustration of the decorating spins. In particular, we have investigated the ground-state and finite-temperature phase diagrams of the mixed spin-1/2 and spin-1 Ising model on a decorated square lattice along with typical thermal dependencies of the spontaneous magnetization and specific heat.

The most interesting finding of the present study concerns with a theoretical prediction of the remarkable frustrated antiferromagnetic phase with a striking coexistence of the antiferromagnetic long-range order of the nodal spins with a disorder of the decorating spins. It has been evidenced that the frustrated antiferromagnetic phase may also exhibit double reentrant phase transitions unlike the classical ferromagnetic or ferrimagnetic phases with a single critical point. Moreover, the mutual interplay between the equilibrium pair exchange interaction, the magnetoelastic coupling, the uniaxial single-ion anisotropy, and the angular frequencies of normal-mode oscillations is at origin of very diverse temperature variations of the total specific heat. In fact, it has been verified that the total specific heat may display variety of temperature dependencies with one or two logarithmic divergences and one or two round maxima of a magnetic origin, which are all superimposed on a standard temperature dependence of the lattice part of the specific heat.

\end{document}